\documentclass[%
reprint,
superscriptaddress,
preprintnumbers,
nofootinbib,
nobibnotes,
 amsmath,amssymb,
 aps, prd
]{revtex4-2}

\usepackage{graphicx}% Include figure files
\usepackage{dcolumn}% Align table columns on decimal point
\usepackage{bm}% bold math
\usepackage{hyperref}% add hypertext capabilities
\usepackage{aas_macros}% Include figure files
\usepackage{amsmath} 
\usepackage{physics}
\usepackage{amssymb}
\usepackage{multirow}
\usepackage{graphicx}
\usepackage[table,xcdraw]{xcolor}

\usepackage{orcidlink}

\graphicspath{{./}{figures/}}

\begin{document}

\preprint{APS/123-QED}

\title{Hyperaccreting Magnetised Neutron Stars inside Rotating Massive Envelopes: Low-Power Jets and Precursor Flares}% Force line breaks with \\

\author{Patrick Chi-Kit \surname{Cheong}~\orcidlink{0000-0003-1449-3363}}
\affiliation{Unaffiliated}

\author{Christopher L. \surname{Fryer}~\orcidlink{0000-0003-2624-0056}}
\affiliation{Center for Nonlinear Studies, Los Alamos National Laboratory, Los Alamos, NM 87545, USA}

\author{David \surname{Radice}~\orcidlink{0000-0001-6982-1008}}
% \thanks{Alfred P.~Sloan Fellow}
\affiliation{Institute for Gravitation and the Cosmos, The Pennsylvania State University, University Park PA 16802, USA}
\affiliation{Department of Physics, The Pennsylvania State University, University Park PA 16802, USA}
\affiliation{Department of Astronomy \& Astrophysics, The Pennsylvania State University, University Park PA 16802, USA}

\date{\today}% It is always \today, today,

\begin{abstract}
The engulfment of a neutron star (NS) by a massive companion initiates a highly dynamic common-envelope (CE) evolution phase.
As the NS spirals into the dense stellar core, it is subjected to hypercritical accretion rates that threaten to rapidly collapse the NS into a black hole (BH).
However, if the infalling envelope possesses sufficient angular momentum and magnetic fields, the NS might survive longer and launch feedback-driving jets.
To investigate this, we perform fully coupled, axisymmetric General Relativistic Magnetohydrodynamic (GRMHD) simulations of hyperaccreting NSs, featuring energy-integrated two-moment neutrino transport and a 13-isotope nuclear reaction network.
We systematically vary the envelope rotation profile and the magnetic field strength of the NS surface ($B_{\rm surf} \sim 5 \times 10^{10} - 5 \times 10^{13}$~G).
In non-magnetised models, we find that envelope rotation naturally forms a centrifugal barrier and a geometrically thick accretion disk, which suppresses the mass accretion rate and lowers the neutrino luminosity; conversely, the intrinsic spin of the NS has a negligible global impact.
In magnetised models, the differential rotation of the accretion flow vigorously amplifies the toroidal magnetic field via the $\Omega$-effect, driving the expansion of magnetic towers.
Crucially, for strong initial surface magnetic fields ($B_{\rm surf} \gtrsim 2.3 \times 10^{13}$~G), the intense magnetic pressure could completely halt the accretion flow at the NS surface and evacuates a low-density polar funnel.
We conclude that while this highly magnetised NS engine successfully delays prompt BH formation and may launch low-power precursor jets (with powers up to ${\sim} 10^{46}~{\rm erg/s}$) capable of generating observable X-ray flares, it lacks the energy budget to unbind the massive envelope, potentially setting the stage for a subsequent BH-driven explosion.
% \begin{description}
% Secondary publications and information retrieval purposes.
% You may use the \texttt{description} environment to structure your abstract;
% use the optional argument of the \verb+\item+ command to give the category of each item. 
% \end{description}
\end{abstract}

\maketitle

%\tableofcontents

\section{\label{sec:intro}Introduction}

Common-envelope evolution (CEE) is a critical, short-lived phase in binary stellar evolution where two stars orbit within a shared envelope \cite{2013A&ARv..21...59I}.
Understanding the dynamics and ultimate outcome of CEE is essential, as it is the primary formation channel for compact binary mergers, the progenitors of gravitational wave events \cite{2000ARA&A..38..113T, 2013A&ARv..21...59I, 2020cee..book.....I, 2023LRCA....9....2R}.

The engulfment of a neutron star (NS) by the envelope of a massive star, typically a red giant or supergiant, can occur through standard CEE \cite{1978ApJ...222..269T, 2013A&ARv..21...59I, 2025Ap&SS.370...11G}, stellar collisions within globular clusters \cite{1992ApJ...389..546B}, or post-supernova interactions \cite{1994ApJ...423L..19L}.
Following the onset of the merger, the NS spirals inward through the extended envelope.
During this inspiral phase, frictional drag and orbital energy deposition unbind a fraction of the outer envelope.
However, once the NS reaches the dense vicinity of the stellar core, the accretion dynamics change qualitatively.
Local densities and temperatures rise sharply, and the mass flux onto the NS can become highly time-dependent and hypercritical, far exceeding the Eddington limit.

Significant effort has been invested to study the global dynamics of such NS--massive star mergers \cite{2015ApJ...798L..19M, 2022MNRAS.514.3212H, 2024ApJ...977..196H, 2024ApJ...971..132E, 2025ApJ...993...61W, 2026MNRAS.551g1542S}.
However, while these global models capture the large-scale dynamics, they often lack the resolution and self-consistent microphysics required to resolve the accretion dynamics of the final merger with the core.

The ultimate fate of this NS-core accretion phase remains a subject of significant debate.
On one hand, the extreme hypercritical accretion rates, coupled with efficient neutrino cooling, suggest that the NS could exceed its maximum mass limit and undergo a quiet collapse to a black hole (BH) on a timescale of minutes or less \cite{1993ApJ...411L..33C, 1995PhR...256...95C, 1996ApJ...459..322C, 1996ApJ...460..801F, 2026PhRvD.114b3018C}.
On the other hand, the infalling envelope material carries significant specific angular momentum.
If this rotation efficiently amplifies magnetic fields, the intense magnetic pressure and centrifugal forces could suppress accretion and launch a jet before or during the NS-core merger.
This jet-feedback mechanism forms the basis of the Common Envelope Jets Supernova (CEJSN) scenario \cite{2019MNRAS.484.4972S, 2025RAA....25b5023S}.
In this CEJSN model, the feedback from the hyperaccreting NS acts as a central engine capable of releasing immense amounts of energy, potentially produce bright supernovae and fundamentally altering the mass ejection process \cite{2015MNRAS.449..288P, 2022MNRAS.514.3212H, 2026arXiv260710267G, 2026arXiv260713023H, 2017MNRAS.470.2929M, 2019MNRAS.482.3646L, 2020MNRAS.497.2057L}.
Consequently, it remains highly uncertain whether the hyperaccreting NS acts as the primary engine for these exceptionally energetic transients, or if it rapidly collapses to a BH which could subsequently power the explosion.

Despite the critical importance of these systems, numerical modeling of a hyperaccreting NS inside a massive envelope is exceptionally challenging.
Capturing the true dynamics requires extreme multi-physics operating in a fully coupled manner: multi-dimensional general relativistic magnetohydrodynamics (GRMHD), self-consistent neutrino transport, and nuclear reaction feedback.
Previous smaller-scale simulations \cite{1996ApJ...460..801F, 2026ApJ...997...88A} have addressed the inner engine but have largely relied on parameterised neutrino cooling and simplified NS surface treatments, and have neglected rotation and magnetic fields.
Consequently, they cannot physically capture the critical boundary-layer heating at the NS surface, which could impact the highly localised convective feedback driven by nuclear burning and magnetorotational effects.
Recent 3D GRMHD simulations \cite{2025ApJ...987...71C} explored the hypercritical accretion of a magnetised disk onto an unmagnetised, non-rotating NS.
However, this work also relied on a simplified treatment of the NS surface and parameterised neutrino cooling, initialised the accretion from a pre-formed disk, and neglected the magnetic field of the NS.

In our previous work \cite{2026PhRvD.114b3018C}, we focused on the late-stage CEE after the NS has completed its orbital inspiral and settled at the center of the giant star envelope (assuming the NS did not collapse into a BH during the inspiral).
Specifically, we presented fully coupled axisymmetric general relativistic hydrodynamic simulations of hypercritical accretion onto an embedded NS, featuring grey two-moment neutrino transport and a 13-isotope $\alpha$-chain nuclear reaction network.
However, that foundational study did not include stellar rotation or magnetic fields.
In this work, we build upon the configuration established in \cite{2026PhRvD.114b3018C} by introducing both envelope rotation and initial magnetic fields.
In this central configuration, the rotational symmetry axis and angular momentum of the surrounding envelope material are aligned with the central NS.
By doing so, we aim to definitively assess how the centrifugal barrier and magnetic tower formation impact the accretion process, determine whether the NS can avoid immediate collapse, and provide diagnostic criteria for the resulting transient precursors.

The paper is organised as follows.
Section~\ref{sec:methods} outlines the methodology and numerical setup.
Section~\ref{sec:results} presents the simulation results, and Section~\ref{sec:discussion} summarises our findings and discusses the broader observational implications.
\section{\label{sec:methods}Methods}

\subsection{\label{sec:id}Initial conditions}

\subsubsection{Neutron Star Models}
The NS models are generated with the \texttt{RNS} code~\citep{1995ApJ...444..306S}, including modifications that enforce conformally flat conditions \cite{1996PhRvD..53.5533C, 2014GReGr..46.1800I, 2021MNRAS.503..850I, 2022MNRAS.510.2948I, 2024PhRvD.110d3015C}. 
We consider three different uniformly rotating configurations, as detailed in Table~\ref{tab:ns_models}, all sharing roughly the same gravitational mass of $1.4~{\rm M_{\odot}}$.
These initial configurations are constructed with a fixed uniform temperature of $0.5~\mathrm{MeV}$ and in neutrinoless $\beta$-equilibrium using the DD2 EoS~\citep{2010NuPhA.837..210H}.

Our selection of NS spin profiles (R0 to R2) is designed to systematically probe the influence of the central engine's rotation.
R0 serves as the static reference ($a_{\rm NS} = 0$), adopted also in \cite{2026PhRvD.114b3018C}. 
Model R1 represents a moderately spinning neutron star ($a_{\rm NS} = 0.20$) with near-spherical geometry ($r_{\rm e}/r_{\rm p} = 0.98$), whereas model R2 represents a rapidly spinning, strongly oblate neutron star ($a_{\rm NS} = 0.69$, $r_{\rm e}/r_{\rm p} = 0.80$) with a order-of-magnitude higher kinetic-to-binding energy ratio ($T/W \approx 5.5\%$).
Although both sit in the millisecond period domain ($P \approx 4.4\text{~ms}$ and $1.5\text{~ms}$, respectively), this choice allows us to span a wide physical range to cleanly isolate the hydrodynamic impact of the NS's oblate geometry and the general relativistic frame-dragging effect on the innermost accretion flow.

\begin{table*}[]
\centering
\resizebox{\textwidth}{!}{%
\begin{tabular}{c|cccccc}
NS Model &
  $\epsilon_{c}/c^2~[10^{14}~\rm{g / cm^{3}}]$ &
  $M_{\rm rest}~{\rm [ M_{\odot}]}$ &
  $\omega~[10^{3}~{\rm rad/s}] $ &
  $a_{\rm NS} $ &
  $\left| T/W \right| [10^{-2}]$ &
  $r_{\rm e} / r_{\rm p}$ \\ \hline
R0 & 6.234 & 1.52803 & 0       & 0    & 0       & 1    \\
R1 & 6.191 & 1.52706 & 1.43648 & 0.20 & 0.51731 & 0.98 \\
R2 & 5.768 & 1.51482 & 4.31666 & 0.69 & 5.54557 & 0.80
\end{tabular}%
}
\caption{    
	Table of the different initial NS models.
	All configurations have a gravitational mass of $M_{\rm g} \approx 1.4~{\rm M_{\odot}}$.
	The quantities $\epsilon_{c}/c^2$, $M_{\rm rest}$, $\omega$, $a_{\rm NS} $, $ \left| T/W \right|$, and $r_{\rm e} / r_{\rm p}$ denote the central mass-energy density, rest mass, angular velocity, spin of the NS, kinetic-to-binding energy ratio, and the equatorial-to-polar radius ratio, respectively.
        }
        \label{tab:ns_models}
\end{table*}

Magnetic fields are added on top of the quasi-equilibrium neutron star profile.
In particular, we superimpose magnetic fields by adding the following vector potential in orthonormal form:
\begin{equation}
	\left(A^{\hat{r}}, A^{\hat{\theta}}, A^{\hat{\phi}}\right) = \frac{r_c^3 }{2\left(r^3+r_c^3\right)}\left(0, 0, B_{\rm c} r \sin\theta \right),
\end{equation}
where the relation between orthonormal-basis components and coordinate-basis components can be found in Appendix~A of \cite{2020CQGra..37n5015C}.
Here, we set $r_c = 8~{\rm km}$.
This vector potential gives us purely poloidal magnetic fields with maximum strength $B_{\rm c}$ at the centre of the NS.
Our models are parameterised by the central magnetic field strength $B_{\rm c}$.
Due to the radial decay of the vector potential, a central field of $B_{\rm c} = 10^{11}$~G corresponds to a polar surface magnetic field of $B_{\rm surf} \approx 4.64 \times 10^{10}$~G at $r_{\rm pole} \approx 10$~km.
Because the magnetic field strength scales linearly in our initial data, our simulation suite ($B_{\rm c} = 10^{11}, 10^{13}, 5\times 10^{13}, \text{and } 10^{14}$~G) spans polar surface fields of $B_{\rm surf} \approx 4.64 \times 10^{10}, 4.64 \times 10^{12}, 2.32 \times 10^{13}, \text{and } 4.64 \times 10^{13}$~G, respectively.
For simplicity, when discussing macroscopic physical scales, we will refer to these approximate surface field values.

\subsubsection{Stellar Model}
The $15~{\rm M_{\odot}}$ extended post-main-sequence star at the terminal-age core helium burning stage is adopted as the stellar progenitor model.
This represents the red supergiant phase, characterised by a large radius and a long lifetime, adopted from our previous study \cite{2026PhRvD.114b3018C}.
The stellar profiles are generated using \texttt{MESA} (Modules for Experiments in Stellar Astrophysics, version 25.12.1)~\cite{Paxton2011, Paxton2013, Paxton2015, Paxton2018, Paxton2019, Jermyn2023}.

\begin{figure}
    \centering
    \includegraphics[width=\columnwidth]{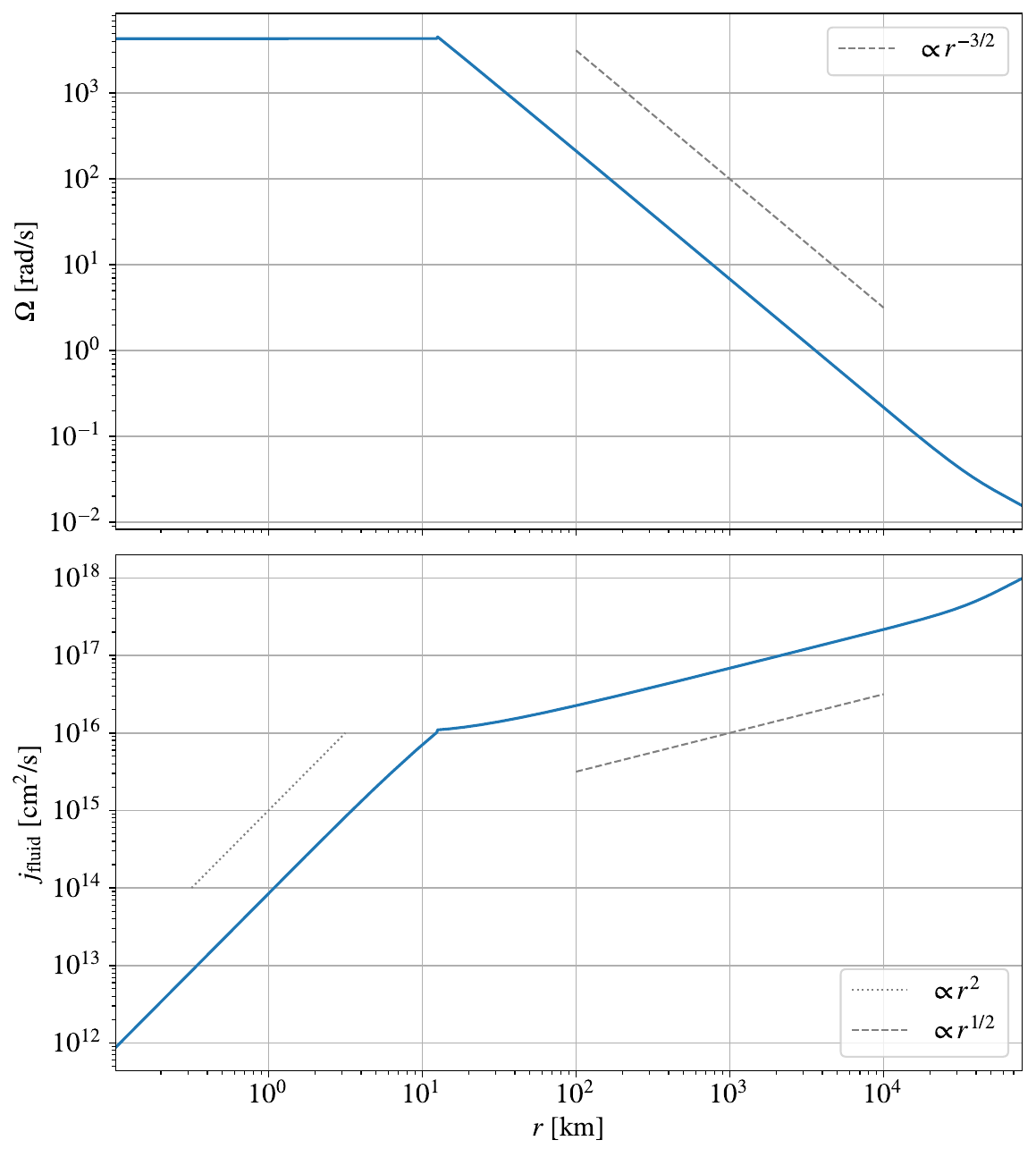}
    \caption{
    Initial equatorial rotational profile of the central engine and surrounding stellar envelope, shown for the representative model with a neutron star dimensionless spin $a_{\rm NS} = 0.69$ and an envelope rotation parameter $\eta = 0.5$. 
    The top panel displays the angular velocity $\Omega$, while the bottom panel shows the specific angular momentum of the fluid $j_{\rm fluid}$, along the equatorial plane. 
    The central NS ($r \lesssim 12$ km) is initialised in uniform solid-body rotation ($\Omega = \text{constant}$), resulting in a steep interior profile of $j \propto r^2$. 
    Beyond the NS surface, the envelope transitions to a differential rotation profile. 
    The dashed grey lines denote the expected analytical scalings for our sub-Keplerian envelope setup ($\Omega \propto r^{-3/2}$ and $j \propto r^{1/2}$). 
    As shown, the outer envelope possesses a substantial angular momentum reservoir ($j \gtrsim 10^{16} \ \mathrm{cm^2/s}$), ensuring that infalling material will eventually circularise to form a centrifugally supported accretion flow at smaller radii.
    }
    \label{fig:id_rot}
\end{figure}

To model the angular momentum of the common envelope, we superimpose a sub-Keplerian angular velocity profile onto the spherically symmetric stellar background.
Specifically, the local angular velocity $\Omega_{\rm env}$ is defined as:
\begin{equation}
    \Omega_{\rm env} = \eta \Omega_{\rm K} \approx \eta \sqrt{ \frac{M_{\rm enc} }{r^3} },
\end{equation}
where $M_{\rm enc}$ is the enclosed mass (dominated in the central regions by the $1.4~{\rm M_{\odot}}$ NS), and $\eta$ is a dimensionless scaling parameter that controls the fraction of Keplerian rotation in the envelope.
At radii characteristic of the inner envelope ($r \sim 10^3 - 10^4~\mathrm{km}$), this yields an initial angular velocity on the order of $\Omega_{\rm env} \sim 1 - 10^{-1}~\mathrm{rad/s}$.
The resulting initial rotational profile is illustrated in Figure~\ref{fig:id_rot} for a representative model, demonstrating the clean transition from a uniformly rotating NS core to a differentially rotating, sub-Keplerian envelope.

This parameterisation provides a straightforward analytical expectation for the accretion dynamics.
Assuming specific angular momentum is conserved during free-fall, a fluid parcel originating at an initial radius $r_0$ will circularise when its centrifugal acceleration balances gravity, occurring at a radius:$r_{\rm circ} = {j^2}/{M_{\rm enc}} = \eta^2 r_0$.  
The choice of $\eta$ therefore strictly dictates the location of the centrifugal barrier.
In this work, we consider $\eta \in \{0.1, 0.3, 0.5\}$ to bracket the distinct physical regimes of hypercritical accretion.
The $\eta = 0.1$ case represents the quasi-radial infall regime, yielding $r_{\rm circ} = 0.01 r_0$.
In this scenario, gas possesses minimal angular momentum, meaning that all material originating within the inner 1200 km plunges almost directly onto the NS surface ($R_{\rm NS} \approx 12$ km) before a disk can form.
Conversely, higher values ($\eta = 0.3$ and $0.5$) ensure that the infalling gas has sufficient specific angular momentum to halt at a larger radius.
For instance, with $\eta = 0.5$, gas circularises at 25\% of its initial radius.
This dynamically forces the formation of an extended, rotationally supported accretion disk and a prominent, stalled accretion shock, producing a robust flow capable of collimating an engine-driven jet well outside the central remnant.

\subsubsection{Simulation Suite}
To systematically isolate the effects of rotation before investigating full magnetorotational dynamics, we perform two distinct series of simulations, comprising a total of 14 models.

First, we consider a suite of six non-magnetised models spanning the parameter space $\eta \in \{0.1, 0.3, 0.5\} \times \{\text{R1}, \text{R2}\}$.
This combination allows us to explore the macroscopic impact of envelope rotation alongside the general relativistic frame-dragging and structural oblateness induced by the intrinsic spin of the NS.

Second, we explore eight magnetorotating models across the parameter space $\eta \in \{0.3, 0.5\} \times B_{\rm c} \in \{10^{11}, 10^{13}, 5\times 10^{13}, 10^{14}\}~\text{G}$.
As discussed in Section~\ref{sec:results}, our purely hydrodynamic runs demonstrate that the intrinsic NS spin profile has a negligible impact on the large-scale evolution of the system.
Therefore, to optimise computational resources and isolate the magnetorotational effects, we fix the NS configuration to the rapidly spinning R2 model ($a_{\rm NS} = 0.69$) for all magnetised configurations.
This grid allows us to cleanly investigate the coupled interplay between moderately and strongly rotating envelopes and a wide range of initial magnetic field strengths.

\subsection{\label{sec:setup}Simulation setup}
The GRMHD simulation setup is essentially the same as our previous work~\cite{2026PhRvD.114b3018C} except that we have magnetic field and rotation here.
In particular, we investigate the deeply embedded core-merger phase under the assumption that the NS has rapidly sunk to the centre of the massive star.
This central configuration is motivated by CEE studies indicating that the NS can survive the dynamical inspiral phase with only modest mass gain due to accretion suppression in the outer envelope~(e.g. \cite{2015ApJ...798L..19M}). 
Although the precise inspiral trajectories and drag timescales remain an active area of research (e.g., \cite{2026MNRAS.551g1542S}), assuming a core-settled state provides a well-defined baseline to isolate the late-stage hyperaccretion engine, where local gas densities and accretion rates are at their highest.
Here, we highlight only the essential parts, and refer readers to \cite{2026PhRvD.114b3018C} for detailed numerical setup.

Our models are evolved with the GR$\nu$MHD code \texttt{Gmunu}~\citep{2020CQGra..37n5015C, 2021MNRAS.508.2279C}, which solves GR$\nu$MHD and Einstein field equations in the conformally flat approximation.
Energy-integrated two-moment 3-species neutrino transport~\citep{2023ApJS..267...38C, 2024ApJ...975..116C}, and nuclear burning~\citep{2026ApJS..284....6C} modules are also activated in a fully coupled manner.
We employ the IMEXCB3a time integrator \citep{2015JCoPh.286..172C}, the Harten-Lax-van Leer (HLL) Riemann solver \citep{harten1983upstream}, and a 3rd-order reconstruction piecewise parabolic method (PPM)~\citep{1984JCoPh..54..174C}. 
The divergence-free condition of the magnetic field is preserved by using staggered-meshed constrained transport \citep{1988ApJ...332..659E, 2022ApJS..261...22C}.
The Lorentz factor $W$ and the magnetisation $\sigma_{\rm mag}$ are capped to be 100 highest.
We allow the NS surface to evolve freely, but freeze the magnetohydrodynamical and radiation variables in the NS interior where $r<8~{\rm km}$ during the evolution. 
The metric is kept fixed entirely.
This fixed metric approximation is well justified as the maximum mass accreted onto the NS across all simulated models is $\Delta M_{\rm acc} \lesssim 10^{-4}~{\rm M_\odot}$ over the entire simulation duration of $t_{\rm sim} \approx 500~{\rm ms}$.
Since $\Delta M_{\rm acc} / M_{\rm NS} \lesssim 10^{-4} \ll 1$, the backreaction of the accreted mass on the spacetime metric is negligible over the timescales considered.

We employ a hybrid equation of state constructed by combining a high-temperature nuclear DD2 EoS~\cite{2010NuPhA.837..210H} with a low-temperature stellar EoS \texttt{helmeos}~\cite{2000ApJS..126..501T}.
At high temperatures (i.e. $T \geq 5.8~\mathrm{GK}$), we assume nuclear statistical equilibrium (NSE) and use the DD2 EoS, which provides thermodynamic quantities for dense, hot matter including contributions from nucleons, nuclei, and leptons.
At lower temperatures (i.e. $T < 5~\mathrm{GK}$), we adopt the \texttt{helmeos} stellar EoS~\cite{2000ApJS..126..501T}, which includes ideal gas ions, photon radiation, and degenerate or relativistic electrons and positrons, in local thermodynamic equilibrium.
In the intermediate region, $5~\mathrm{GK} \leq T < 5.8~\mathrm{GK}$, all thermodynamic quantities are obtained via linear interpolation between the two EoSs.
NSE is enforced in the intermediate region.

The neutrino microphysics is provided by \texttt{NuLib}~\citep{2015ApJS..219...24O}.
The table includes reaction rates for the charged current reactions $p + e^- \leftrightarrow n+\nu_e$ and $n+e^+ \leftrightarrow p+\bar\nu_e$;
scattering of neutrinos on protons, neutrons, $\alpha$-particles and heavy nuclei; $e^+e^- \leftrightarrow \nu\bar \nu$ and Bremsstrahlung for the heavy-lepton neutrinos only. 
Neutrino-matter interactions are enabled only in regions where neutrinos are expected to interact significantly with matter.
Specifically, the neutrino-matter coupling is activated when $\rho \geq 10^5~{\rm g \cdot cm^{-3}} $ and $T \geq 0.431~{\rm MeV}$, which are typical thresholds for appreciable neutrino optical depths in dense accretion flows.
Outside this regime, neutrinos are treated as free streaming, and only their advection along characteristics is tracked without energy or momentum feedback onto the hydrodynamics.

Fully coupled nuclear burning \citep{2026ApJS..284....6C} is employed.
We adopt the standard 13-isotope $\alpha$-chain network {\tt aprox13} \citep{1999ApJS..124..241T, 2000ApJS..129..377T}, consisting of $\rm ^4He$, $\rm ^{12}C$, $\rm ^{16}O$, $\rm ^{20}Ne$, $\rm ^{24}Mg$, $\rm ^{28}Si$, $\rm ^{32}S$, $\rm ^{36}Ar$, $\rm ^{40}Ca$, $\rm ^{44}Ti$, $\rm ^{48}Cr$, $\rm ^{52}Fe$, and $\rm ^{56}Ni$.
Free neutrons and protons are also advected as passive species but do not participate in any nuclear reactions within this network.
Nuclear reactions are activated only within the temperature range $10^6~{\rm K} \leq T \leq 5\times10^{9}~{\rm K}$.

The axisymmetric simulations are performed in isotropic cylindrical coordinates $(R_{\rm iso}, z_{\rm iso})$ over a computational domain extending across $R_{\rm iso} \in [0, 8 \times 10^4]~{\rm km}$ and $z_{\rm iso} \in [-8\times 10^4, 8 \times 10^4]~{\rm km}$.
The base grid consists of $N_R = 128$ and $N_z = 256$ cells, and we employ up to $l_{\max}=13$ adaptive mesh refinement (AMR) levels, yielding a finest grid spacing of $\Delta R_{\rm iso} = \Delta z_{\rm iso} \approx 153~{\rm m}$.
Following the methodology established in \cite{2026PhRvD.114b3018C}, our mesh refinement strategy utilises a dual-layered approach that couples fixed spatial nesting with dynamic feature tracking.
This framework incorporates a radially dependent refinement level limiter supplemented by an adaptive shock-front tracking algorithm.

For the purely hydrodynamical simulations, we enforce equatorial plane symmetry across $z=0$, which reduces the active computational domain by half.
Conversely, for the magnetised configurations, the full domain is simulated without symmetry assumptions.
For these magnetised runs, we dynamically monitor both the magnetorotational instability (MRI) quality factor, $Q_{\rm MRI}$, and the fluid magnetisation parameter, $\sigma_{\rm mag} \equiv b^2/\rho$, to better resolve magnetic field amplification in intermediate-to-high density regions.
Additional grid refinement is actively triggered whenever the flows satisfy $Q_{\rm MRI} < 8$ within the density regime $\rho \geq 10^6~{\rm g/cm^3}$, or whenever $\sigma_{\rm mag} > 1$.
We note that although a low $Q_{\rm MRI}$ value prompts local refinement, the maximum achievable resolution remains strictly capped as a function of radius according to the prescription detailed in \cite{2026PhRvD.114b3018C}.
Consequently, as discussed further in Section~\ref{sec:results}, our global grid resolution is ultimately insufficient to fully resolve the MRI throughout the entire domain.

% The MRI quality factor is computed as
% \begin{equation}
% 	Q^z = \frac{\lambda_{\rm MRI}^z}{\psi^2 \Delta z},
% \end{equation}
% where
% \begin{equation}\label{eq:lambda_mri}
% 	\lambda_{\rm MRI} = \frac{2 \pi v_{\rm A}^z}{\Omega} \approx \frac{2\pi \left| B_{\rm pol} \right| }{\Omega \sqrt{\rho}}.
% \end{equation}
% Since the $Q$-factor is used for refinement during the simulation and not necessary to be exact, we used the approximated $\lambda_{\rm MRI}$ for simplicity.

\subsection{\label{sec:diagnostics}Diagnostics}

We define the mass accretion rate as
\begin{equation}
    \dot{M}_{\rm acc} = - \oint D \hat{v}^r \sqrt{\gamma} \dd \theta \dd \phi,
\end{equation}
where $\hat{v}^i = \alpha v^i - \beta^i$, $\alpha$ and $\beta^i$ are the lapse function and shift vector of the spacetime.
$D = W\rho$ is the conserved rest-mass density, and $\gamma$ is the determinant of the 3-metric. 

The energy flux is defined as
\begin{equation}\label{eq:L_eng}
    {L}_{\rm energy} = \oint \left[  \tau \hat{v}^r + \alpha p^* v^r - \alpha^2 b^0 B^r / W \right] \sqrt{\gamma} \dd{\theta} \dd{\phi},
\end{equation}
where
\begin{align}
    {\tau} &= \rho h^* W^2 - p^* - ( \alpha b^0 )^2 - D,\\
    h^* &= 1 + \left( \varepsilon - \varepsilon_0 \right) + (p + b^2) / \rho, \\
    p^* &= p + b^2/2. 
\end{align}
Here, $p$ is the fluid pressure, $b^{\mu}$ is the fluid-frame magnetic field 4-vector, and $\varepsilon$ is the fluid specific internal energy.
The quantity $\varepsilon_0$ is a reference zero-point, evaluated at the same rest-mass density and composition but at zero temperature~\citep{2020PhRvD.102l3015B}.
To isolate the power of the magnetically dominated polar funnel, we further define the jet power, $L_{\rm jet}$, using Equation~\eqref{eq:L_eng} but restricting the integration solely to regions where the magnetisation exceeds unity ($\sigma_{\rm mag} = b^2/\rho > 1$).

The diagnostic unbound matter and the explosion energy are given by
\begin{align}
	M_{\rm b, unbound} &= \int_{\rm unbound} \left[ W \rho \right]  \sqrt{\gamma}  \dd{x}^3, \\
	E_\mathrm{exp} &= \int_{\rm unbound} \left[ W \rho \left( -h^* u_t - 1 \right) \right]  \sqrt{\gamma}  \dd{x}^3,
\end{align}
where the unbound condition is defined as 
\begin{align}
	-h^* u_t - 1 &> 0, \\
	v^r &> 0.
\end{align}
Here, $u_t = W(- \alpha + \beta_i v^i )$ is the covariant time component of the four-velocity.

The poloidal and toroidal magnetic field energies are given by
\begin{align}
	{E}_{B_{\rm pol}} &= \int \frac{1}{2} \left( B^R B_R + B^z B_z \right)\sqrt{\gamma} \dd{x}^3, \\
	{E}_{B_{\rm tor}} &= \int \frac{1}{2} \left( B^\phi B_\phi \right)\sqrt{\gamma} \dd{x}^3.
\end{align}

To assess rotational support, we compare the specific angular momentum of the gas against the Keplerian specific angular momentum of a cold disk in Kerr spacetime, $j_{\rm K}$, expressed in Boyer-Lindquist coordinates:
\begin{equation}\label{eq:j_K}
    j_{\rm K} = \frac{ \sqrt{M r_{\rm BL}} \left(  r_{\rm BL}^2 - 2a\sqrt{M r_{\rm BL}} + a^2\right) }{ r_{\rm BL} \sqrt{ r_{\rm BL}^2 -3M r_{\rm BL} + 2a\sqrt{M r_{\rm BL}}}},
\end{equation}
where $M = M_{\rm NS}$ is the NS mass and $a = a_{\rm NS}$ is the dimensionless spin parameter.
Because this expression depends on the Boyer-Lindquist radial coordinate $r_{\rm BL}$, we transform our isotropic coordinate $r_{\rm iso}$ via
\begin{equation}
    r_{\rm BL} = r_{\rm iso} + M + \frac{M^2 - a^2 }{ 4 r_{\rm iso}}.
\end{equation}
Because the accretion disks formed in our simulations are geometrically thick and supported by thermal pressure in addition to rotation, we utilise a sub-Keplerian threshold to identify the disk material.
Specifically, we define the disk region as fluid cells satisfying 
\begin{equation}
\begin{aligned}\label{eq:disk}
	j_{\rm fluid}/j_{\rm K} & \geq 0.6, \\
	10^{4}~{\rm g/cm^3} \leq &\rho \leq 10^{11}~{\rm g/cm^3}.
\end{aligned}
\end{equation}
The density cutoff of is to exclude the NS and the low-density ambient environment.

We note that our conformally flat condition (CFC) metric cannot exactly reproduce the Kerr spacetime.
Furthermore, evaluating the specific angular momentum for circular geodesics directly from the CFC metric would require computing spatial derivatives of the numerical metric components (i.e., the lapse $\alpha$, shift $\beta^\phi$, and conformal factor $\psi$).
To avoid introducing numerical noise from finite differencing in AMR grid for what is ultimately an approximate diagnostic threshold ($j_{\rm fluid}/j_{\rm K} \geq 0.6$) for a highly dynamic disk, we employ the analytical Kerr expression in \eqref{eq:j_K}.
This provides a smooth, convenient reference scale to estimate the disk boundaries.

The total neutrino luminosity $L_\nu$ observed by a distant observer is calculated as
\begin{equation}
    L_{\nu} = \oint \left( \alpha F^r_\nu - \beta^r E_\nu \right) \sqrt{\gamma} \, \dd \theta \dd \phi.
\end{equation}

Finally, the angle-averaged shock radius is defined as
\begin{equation}
    \langle r_{\rm shock} \rangle = \frac{\int_0^{\pi} r_{\rm shock}(\theta) \sin\theta \, \dd \theta}{\int_0^{\pi} \sin\theta \, \dd \theta},
\end{equation}
where $r_{\rm shock}(\theta)$ is the physical areal radius (i.e., $r\psi^2$) of the shock along a given polar angle $\theta$. 
To locate shock radius, we employ a kinematic shock-detection algorithm. 
Searching radially inward from the outer boundary along a given angle $\theta$, the shock front is identified as the outermost location where the infalling gas ($v_r < 0$) undergoes both physical compression ($\Delta \rho > 0$) and a sudden deceleration, defined by a fractional relative velocity jump of $\Delta v_r / |v_r| > 0.1$ across adjacent grid cells. 
We found this kinematic approach provides a stable estimate of the global shock morphology for our specific setup.

\section{\label{sec:results}Results}

\subsection{\label{sec:rot}Non-magnetised cases}

In all non-magnetised models, the envelope material accretes onto the central NS, shocking at its surface and forming an accretion shock, consistent with our previous findings for non-rotating systems \cite{2026PhRvD.114b3018C}.
However, the introduction of angular momentum in the NS and the surrounding envelope fundamentally alters the subsequent evolution.
As we demonstrate below, the rotation of the envelope dictates the strength of the centrifugal barrier and dominates the global dynamics, whereas the intrinsic spin of the NS (and hence the frame dragging) plays only a marginal role.

\begin{figure*}
	\centering
	\includegraphics[width=\textwidth, angle=0]{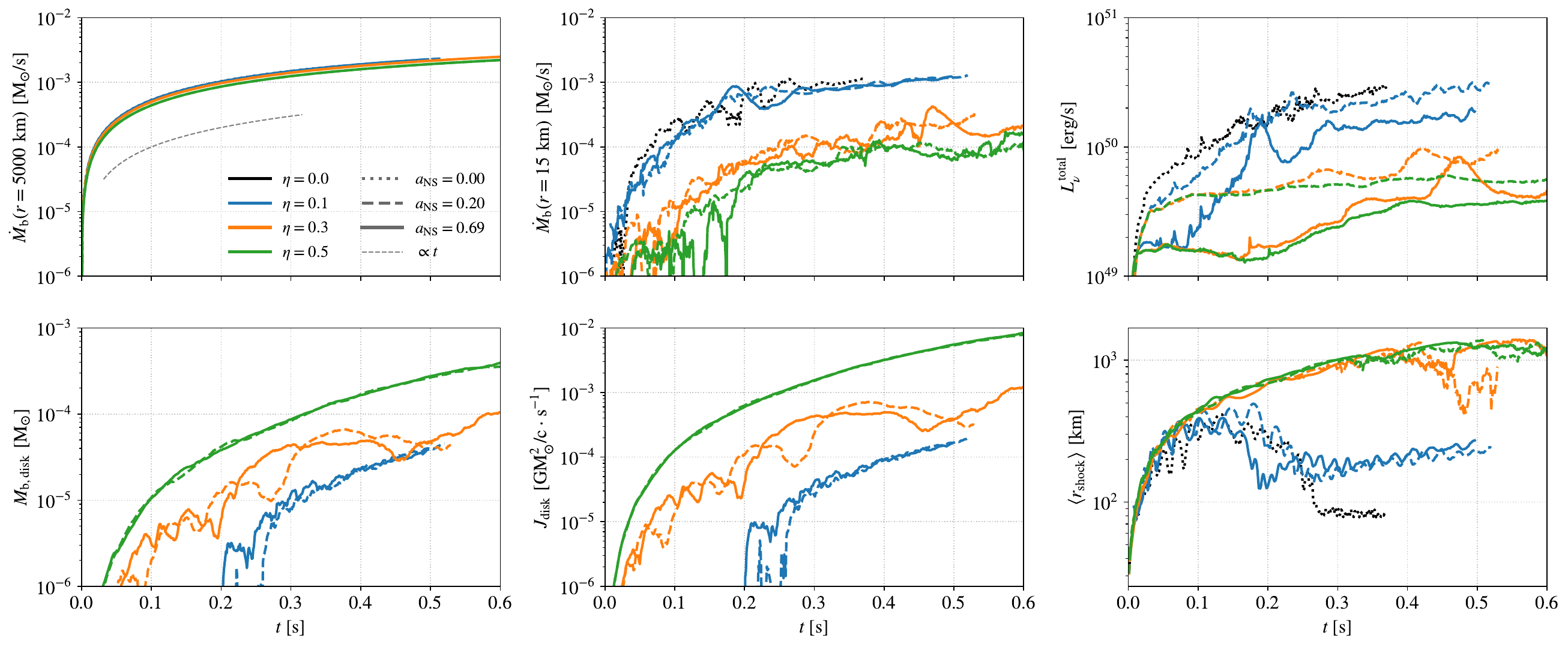}
	\caption{
		Mass accretion rate at 5000~km (\emph{top left}) and 15~km (\emph{top mid}), total neutrino luminosity (\emph{top right}), disk mass (\emph{bottom left}) and angular momentum (\emph{bottom middle}), and the averaged shock radius (\emph{bottom right}) as functions of time of models with different rotations.
	}
	\label{fig:rot_mdot_vs_t}
\end{figure*}

Figure~\ref{fig:rot_mdot_vs_t} summarises the time evolution of all non-magnetised models.
In the pre-shock region (e.g., at $r=5000~{\rm km}$, top-left panel of Figure~\ref{fig:rot_mdot_vs_t}), the mass accretion rate $\dot{M}_{\rm acc}$ is primarily governed by the gravitational potential of the central object and remains virtually identical across all models.
Consistent with the analytical expectations outlined in \cite{2026PhRvD.114b3018C}, $\dot{M}_{\rm acc}$ initially increases approximately linearly with time, reflecting the free-fall collapse of the initially uniform envelope, before eventually reaching a saturated state.

\begin{figure*}
	\centering
	\includegraphics[width=\textwidth, angle=0]{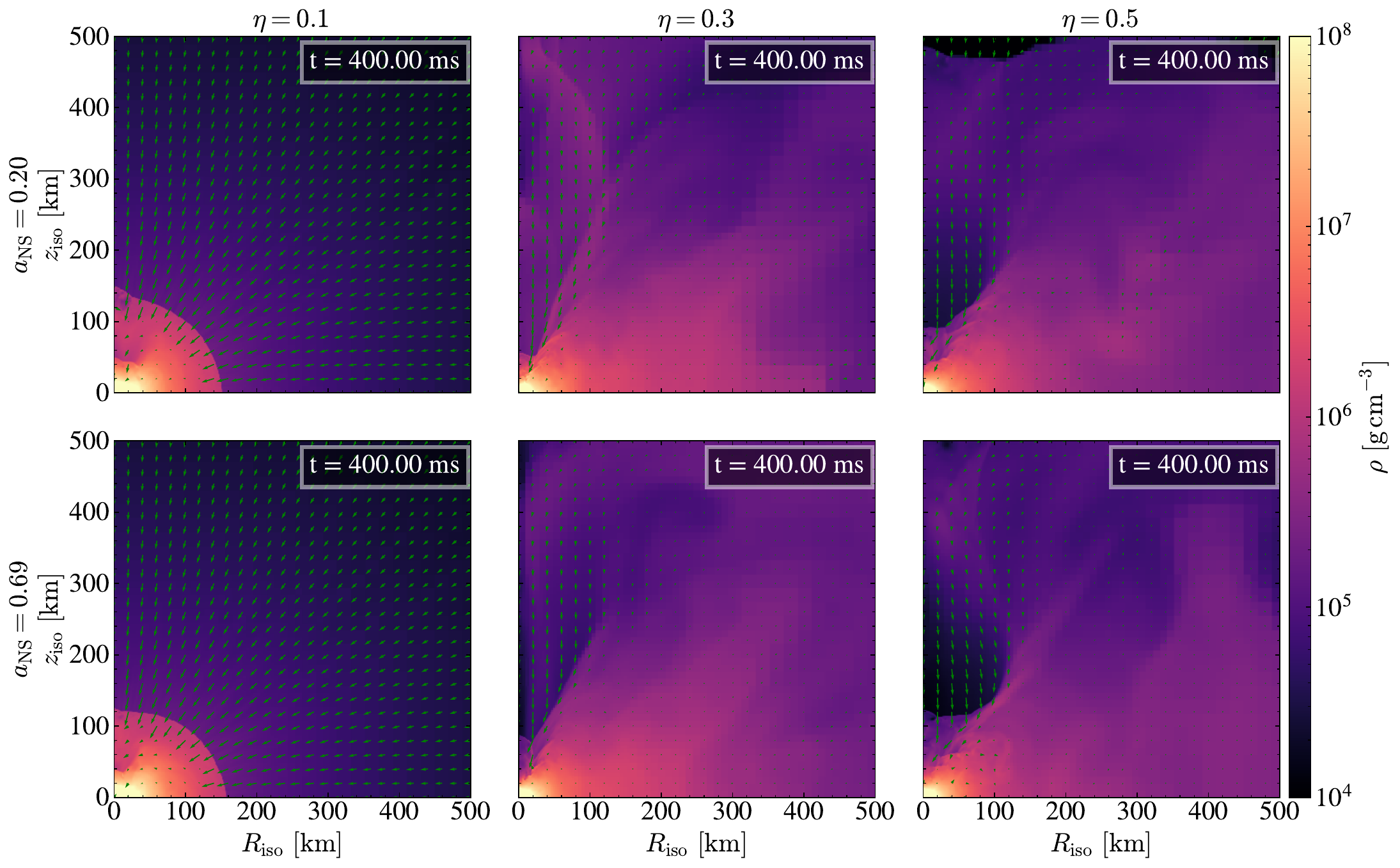}
	\caption{
		Rest mass density $\rho$ profiles for non-magnetised models at $t=400~{\rm ms}$.
		Green arrows indicate the velocity vectors.
	}
	\label{fig:rot_rho}
\end{figure*}

\begin{figure*}
	\centering
	\includegraphics[width=\textwidth, angle=0]{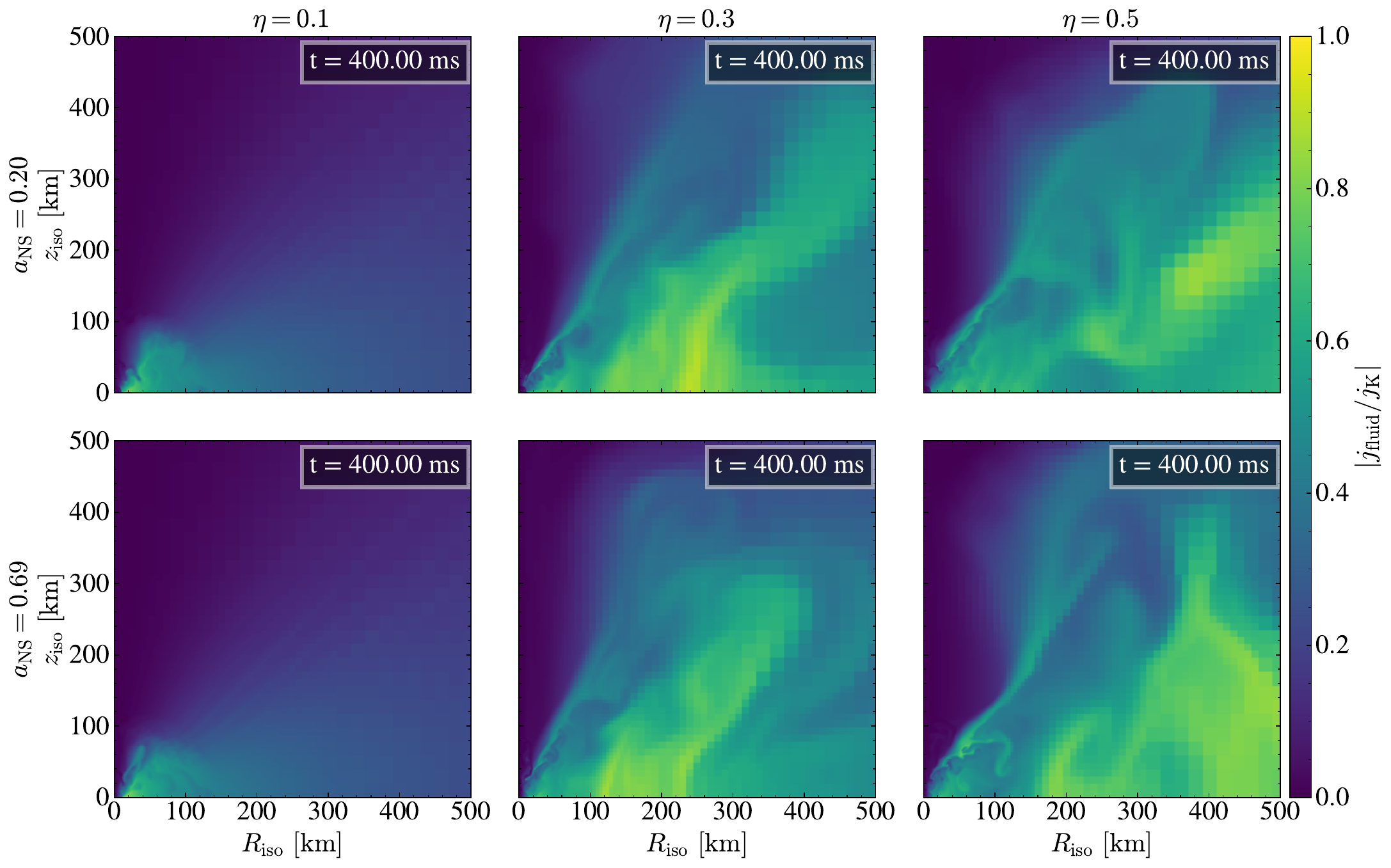}
	\caption{
		Specific angular momentum ratio $j/j_{\rm K}$ profiles for non-magnetised models at $t=400~{\rm ms}$.
		% The red contour in the bottom panels denotes $j/j_{\rm K} = 0.6$ to highlight the disk region.
	}
	\label{fig:rot_j}
\end{figure*}

While the infall in the pre-shock region is quasi-spherical, the post-shock dynamics exhibit stark deviations from spherical symmetry.
The specific angular momentum of the infalling material provides centrifugal support against the gravitational pull, causing the post-shock region to become torus like (see Figure~\ref{fig:rot_rho}) and pushing the shock front to larger radii (Figure~\ref{fig:rot_mdot_vs_t}, bottom-right panel).
This centrifugal barrier promotes the formation of a rotationally supported accretion disk within the post-shock regions (see Figure~\ref{fig:rot_j}).
Consequently, the mass accretion rate onto the NS is significantly suppressed, particularly along the equatorial plane.
This suppression is clearly visible in the accretion rate evaluated at $r=15~{\rm km}$, just above the NS surface (Figure~\ref{fig:rot_mdot_vs_t}, top-middle panel): models with stronger envelope rotation exhibit correspondingly lower inner accretion rates.
This reduction in the accretion rate directly translates to a lower rate of gravitational binding energy dissipation, leading to a noticeable suppression of the total neutrino luminosity, as shown in the top-right panel of Figure~\ref{fig:rot_mdot_vs_t}.

Finally, we evaluate the accumulation of mass and angular momentum within the newly formed disk, where the disk is defined as equation~\eqref{eq:disk}.
As illustrated in the bottom-left and bottom-middle panels of Figure~\ref{fig:rot_mdot_vs_t}, stronger initial envelope rotation leads to a more rapid accumulation of both disk mass and angular momentum.
The intrinsic spin of the NS has negligible influence on the disk formation process for the extreme cases (highest and lowest envelope rotation), and only a very minor, albeit observable, effect in the intermediate rotation case.
It is important to note that the disks formed during this early phase are geometrically thick, supported not only by rotation but also by immense thermal pressure.
Because the disk mass and angular momentum are still growing at the end of our simulation window due to ongoing accretion, we anticipate that more massive disks will eventually form.
Furthermore, as sustained neutrino cooling gradually radiates away the thermal energy on longer timescales, we expect these structures to deflate and transition into thinner, more rotationally supported disks.

\subsection{\label{sec:mag}Magnetised cases}

The introduction of magnetic fields introduces additional complexity to the accretion dynamics.
As demonstrated in Section~\ref{sec:rot}, the spin of the central NS exerts a negligible influence on the global hydrodynamic evolution. 
While the NS spin is expected to play a bigger role in the magnetised regime, exploring this additional dimension of the parameter space is beyond the scope of this work.
Consequently, for all magnetised models discussed below, we fix the dimensionless NS spin to $a_{\rm NS} = 0.69$.
This allows us to isolate and systematically investigate the interplay between the rotation of the stellar envelope (parameterised by $\eta$) and the initial magnetic field configuration (governed by $B_{\rm surf}$).

\begin{figure*}
	\centering
	\includegraphics[width=\textwidth, angle=0]{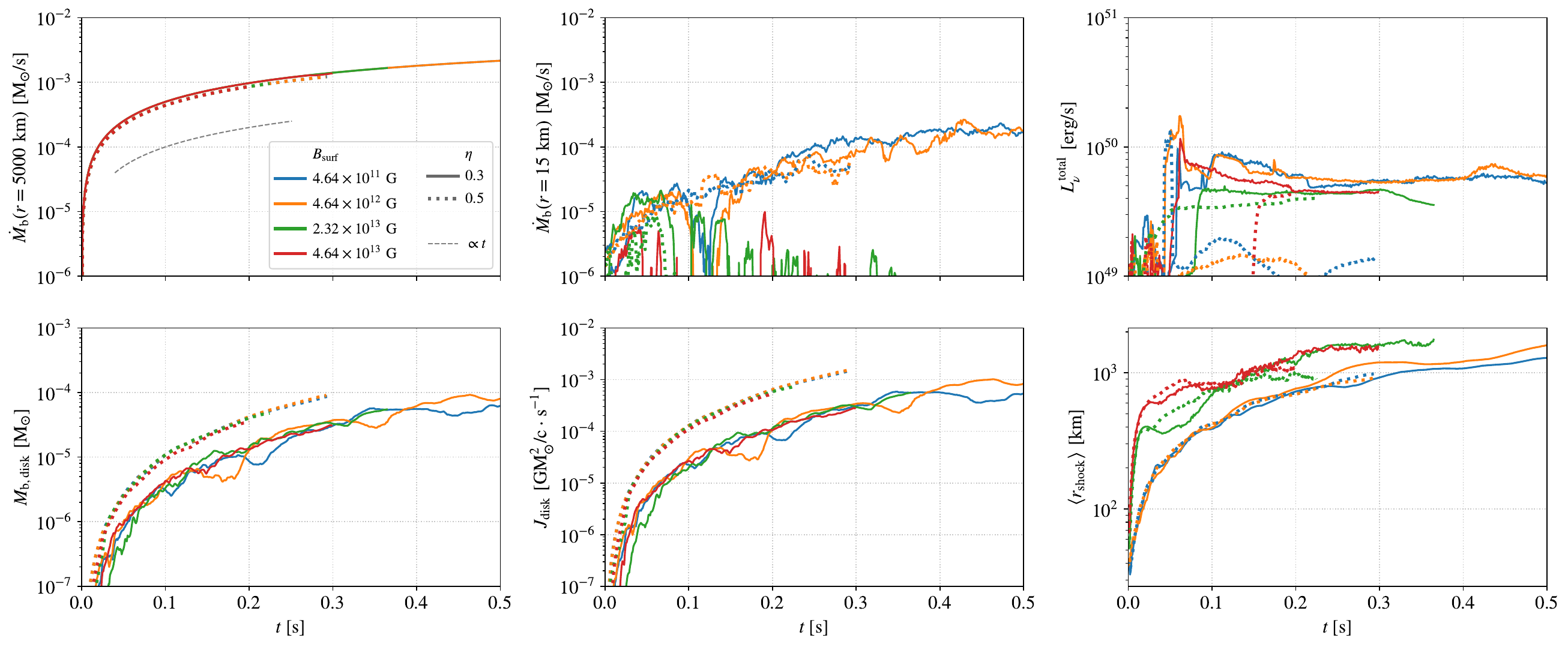}
	\caption{
		Mass accretion rate at 5000~km (\emph{top left}) and 15~km (\emph{top mid}), total neutrino luminosity (\emph{top right}), disk mass (\emph{bottom left}) and angular momentum (\emph{bottom middle}), and the averaged shock radius (\emph{bottom right}) as functions of time of different magnetised models.
	}
	\label{fig:mag_mdot_vs_t}
\end{figure*}

Figure~\ref{fig:mag_mdot_vs_t} summarises the global temporal evolution of the magnetised models.
At large scales ($r = 5000\text{~km}$), the mass accretion rate behaves identically to the non-magnetised baselines, characterised by the quasi-spherical, free-fall collapse of the initially uniform envelope.
Similarly, the total disk mass and accumulated angular momentum are predominantly governed by the initial rotation profile of the envelope, aligning with the trends observed in the purely hydrodynamic cases.

\begin{figure*}
    \centering
    \includegraphics[width=\textwidth, angle=0]{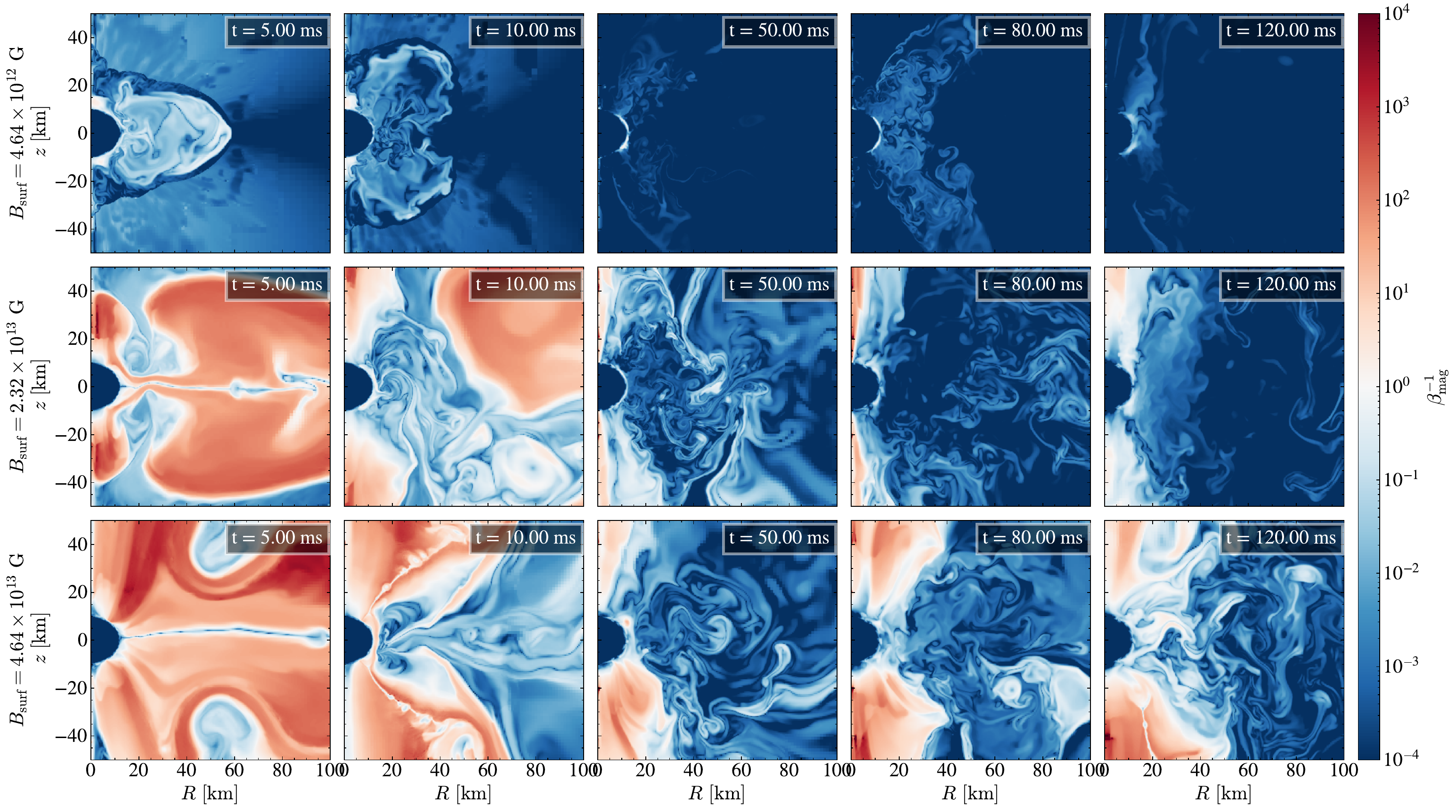}
    \caption{
        Inverse plasma beta ($\beta^{-1}_{\rm mag}$) profiles for magnetised models with $\eta = 0.3$ at various times.
	The $B_{\rm surf}=4.64 \times 10^{10}$~G model is excluded, as it remains strictly fluid-dominated ($\beta^{-1}_{\rm mag} \leq 10^{-4}$).
    }
    \label{fig:mag_beta_mag_time_series}
\end{figure*}

In contrast, the localised dynamics within the central engine—specifically regions in the immediate vicinity of the NS—deviate substantially from the non-magnetised models.
Near the NS surface ($r = 15\text{~km}$), stronger initial magnetic fields result in a more pronounced suppression of the local mass accretion rate (see the regions where $\beta_{\rm mag} \equiv p_{\rm mag} / p_{\rm gas} \gtrsim 1$ in Fig.~\ref{fig:mag_beta_mag_time_series}, indicating magnetic-pressure dominance).
Interestingly, for the $B_{\rm surf} \geq 2.32 \times 10^{13}~{\rm G}$ cases, the accretion onto NS can be halted completely.
This reduction in the inner accretion flow directly impacts the thermal properties of the plasma, subsequently altering the total neutrino luminosity ($L_\nu$), which decreases systematically with increasing magnetic pressure.
Furthermore, this enhanced magnetic pressure provides additional radial support, driving the time-averaged shock radius further outward compared to the purely hydrodynamic cases.

\begin{figure*}
	\centering
	\includegraphics[width=\textwidth, angle=0]{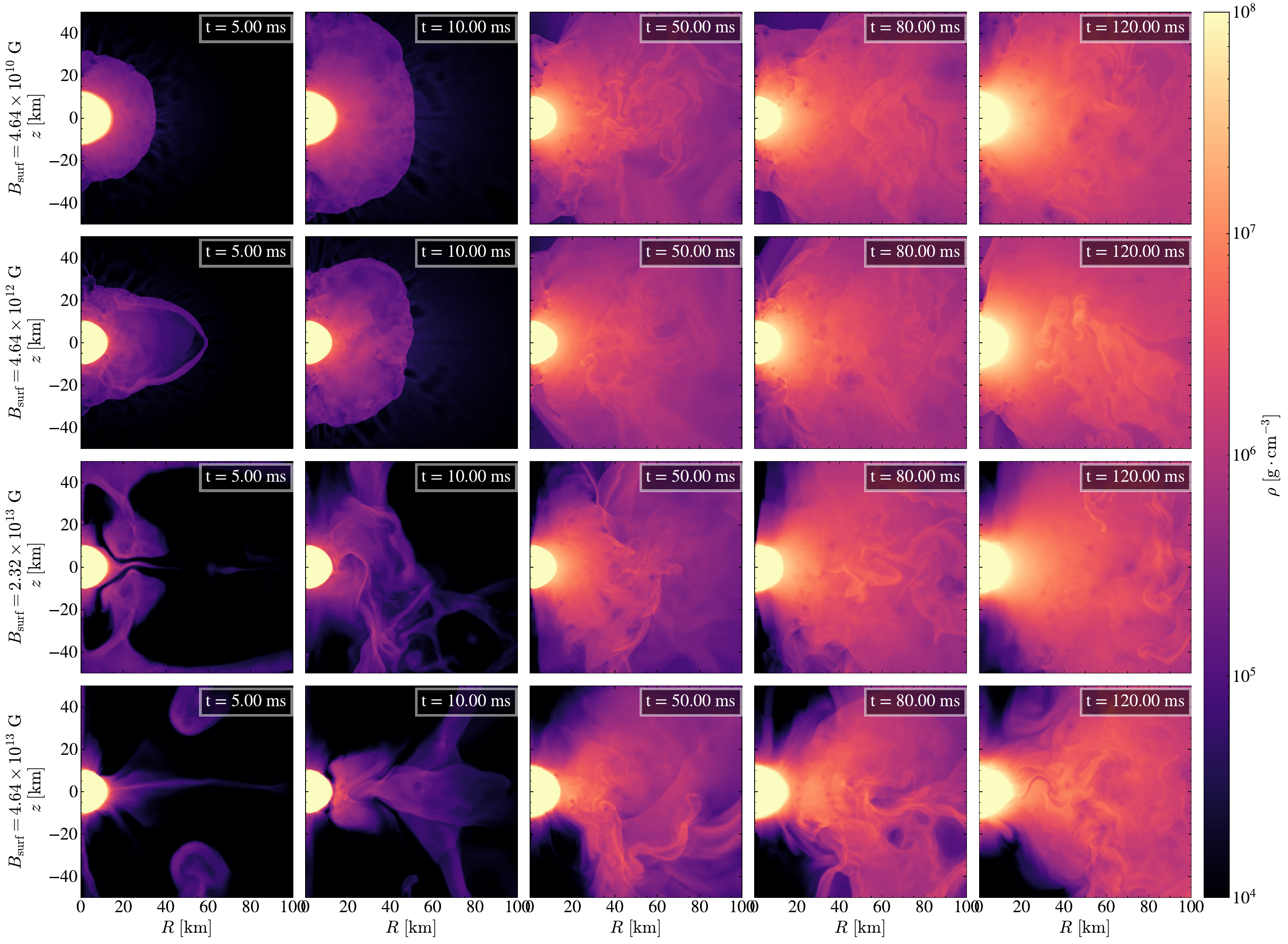}
	\caption{
		Rest mass density ($\rho$) profiles for all magnetised models with $\eta = 0.3$ at various times.
	}
	\label{fig:mag_rho_time_series}
\end{figure*}

While the large-scale envelope rotation dictates the macroscopic mass and angular momentum budget of the disk, the spatial morphology of the accretion flow is highly sensitive to the magnetic field strength.
Figure~\ref{fig:mag_rho_time_series} displays the time evolution of the rest-mass density ($\rho$) profiles for all $\eta = 0.3$ models across varying levels of magnetisation.
For the weakly magnetised case ($B_{\rm surf} = 4.64 \times 10^{10}\text{~G}$), magnetic forces exert a negligible effect on the fluid dynamics; the accretion shock remains initially quasi-spherical before settling into a thick, geometrically extended torus at later times, mimicking the purely rotating models.
However, as the initial field strength is increased, a distinct polar structure emerges.
Enhanced magnetic pressure evacuates matter from the rotation axis, carving out a low-density polar funnel and channeling the primary accretion flow into a dense, equatorially concentrated structure.

\begin{figure}
	\centering
	\includegraphics[width=\columnwidth, angle=0]{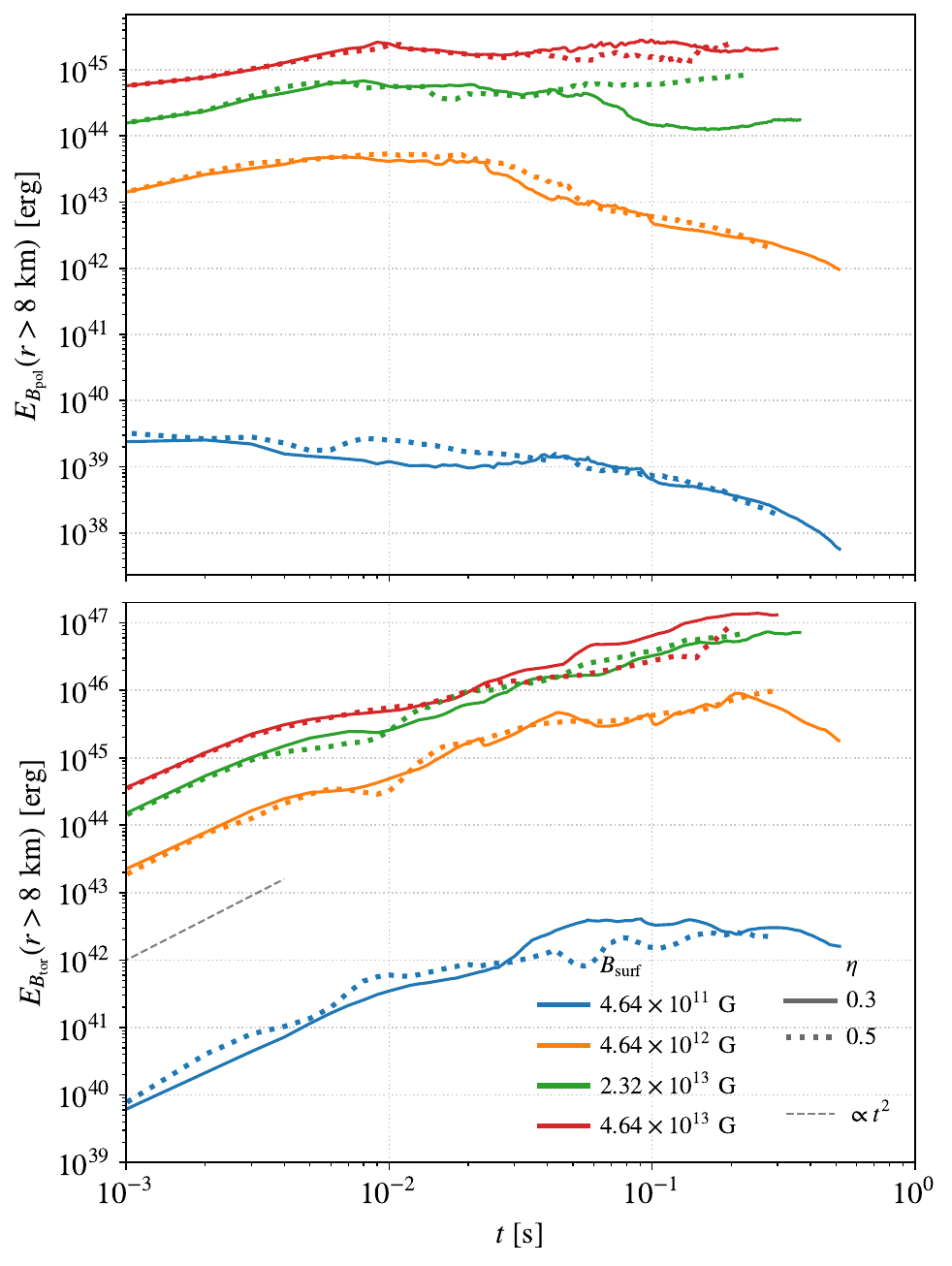}
	\caption{
		Time evolutions of the poloidal and toroidal magnetic field energies outside the NS core ($r>8~{\rm km}$) of all models.
	}
	\label{fig:mag_emag_vs_t}
\end{figure}

Figure~\ref{fig:mag_emag_vs_t} tracks the temporal evolution of the volume-integrated poloidal ($E_{B_{\rm pol}}$) and toroidal ($E_{B_{\rm tor}}$) magnetic energies computed outside the frozen NS core ($r > 8\text{~km}$).
The poloidal magnetic energy remains relatively stagnant and eventually undergoes a gradual decay.
This behaviour is a direct consequence of both our geometric and numerical constraints.
First, because our simulations are restricted to 2D axisymmetry, Cowling's anti-dynamo theorem strictly prohibits the regeneration of poloidal flux via a non-axisymmetric, turbulent $\alpha$-effect dynamo.
Second, while the differential rotation of the accretion flow satisfies the conditions for the MRI, our global grid resolution is capped below the threshold required to fully capture the fastest-growing linear MRI modes.
Consequently, a sustained, exponential growth of the global poloidal field is absent in Figure~\ref{fig:mag_emag_vs_t}.

\begin{figure*}
	\centering
	\includegraphics[width=\textwidth, angle=0]{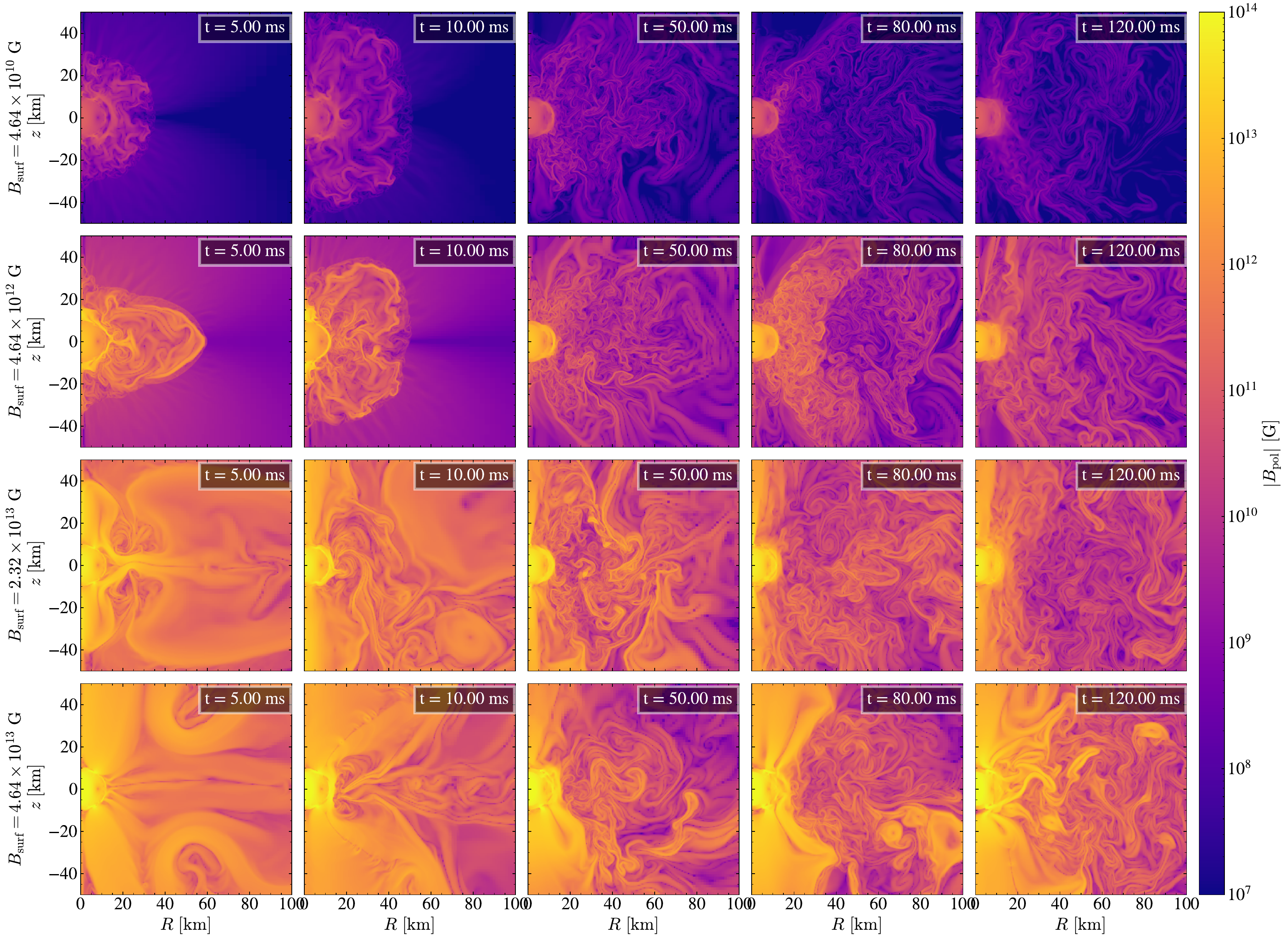}
	\caption{
		Poloidal magnetic field strength ($\left| B_{\rm pol} \right| $) profiles for all magnetised models with $\eta = 0.3$ at various times.
	}
	\label{fig:mag_Bpol_time_series}
\end{figure*}

Nevertheless, the longer-wavelength, slower-growing modes of the MRI spectrum are partially resolved in our high-$B_{\rm surf}$ runs.
Because the characteristic MRI wavelength scales as $\lambda_{\rm MRI} \propto v_A/\Omega$, its detailed value depends on both the magnetic field strength and the local matter state.
Thus, models with stronger fields tend to possess larger, more easily resolvable characteristic wavelengths for a given grid spacing.
This resolution effect manifests as a transient growth phase in $E_{B_{\rm pol}}$ during the early stages of the simulation ($t \lesssim 10\text{~ms}$).
These partially captured modes stir the plasma, generating the localised, filamentary ``wrinkling'' of the poloidal field lines visible in the spatial profiles of $|B_{\rm pol}|$ (Figure~\ref{fig:mag_Bpol_time_series}).
Notably, because the angular velocities ($\Omega$) of the $\eta=0.3$ and $0.5$ envelopes are of the same order of magnitude, the characteristic MRI wavelengths remain comparable in order of magnitude between the two rotation cases.
Consequently, for a given $B_{\rm surf}$, the transient amplification of $E_{B_{\rm pol}}$ is largely insensitive to the $\eta$ values considered, resulting in highly similar early-time energetic trajectories.
Once these local perturbations saturate, the poloidal energy exhibits a slow, monotonic decline driven by the advection of poloidal flux across the NS core and numerical reconnection at the grid scale.
For the weakest field case ($B_{\rm surf} = 4.64 \times 10^{10}\text{~G}$), $\lambda_{\rm MRI}$ is strictly sub-grid scale; as a result, no transient amplification is captured, and $E_{B_{\rm pol}}$ undergoes immediate numerical decay.

\begin{figure*}
	\centering
	\includegraphics[width=\textwidth, angle=0]{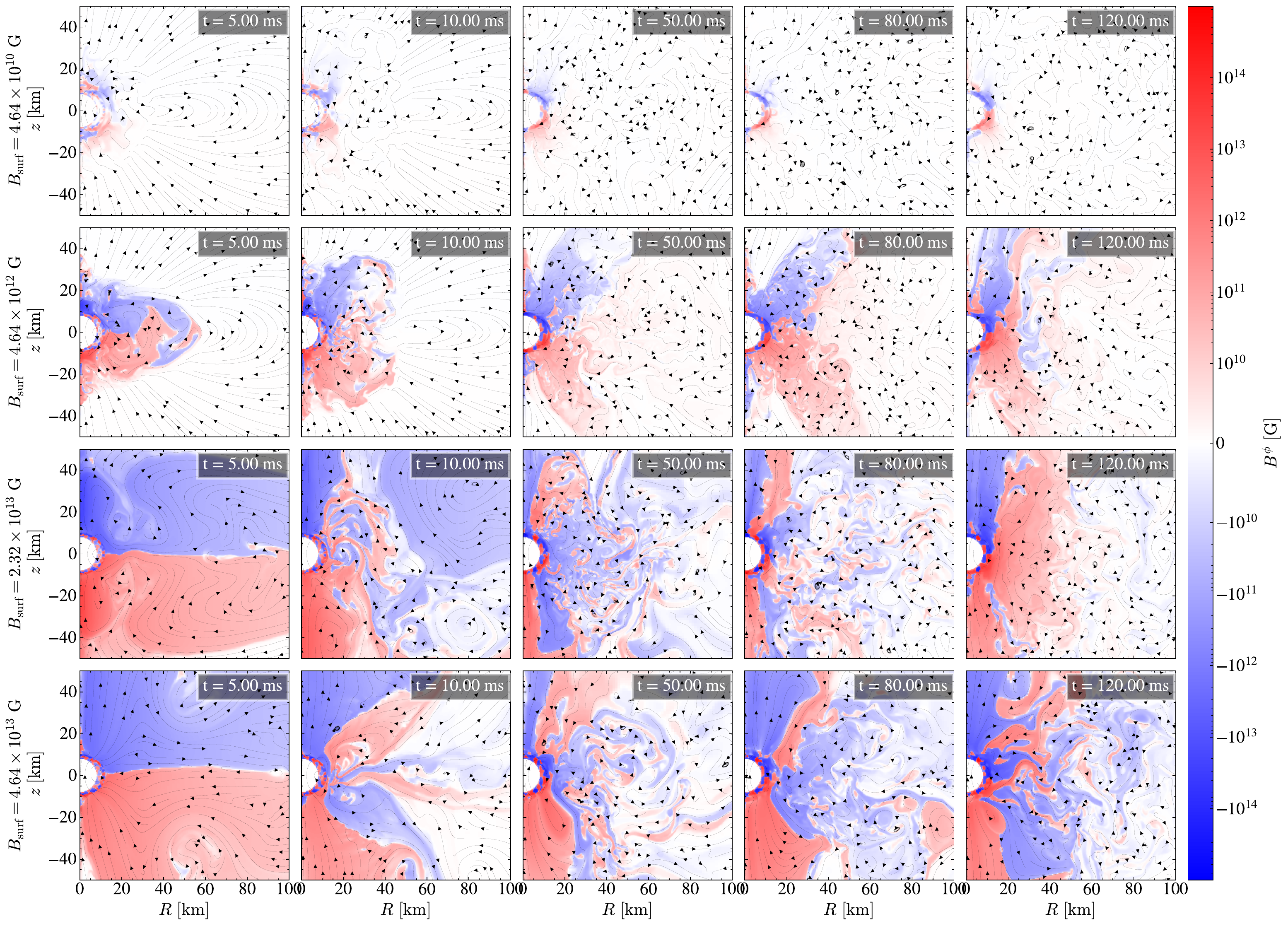}
	\caption{
		Toroidal magnetic field ($B^{\phi}$) profiles for all magnetised models with $\eta = 0.3$ at various times.
		Black streamlines with arrows indicate the poloidal magnetic fields.
	}
	\label{fig:mag_Bphi_time_series}
\end{figure*}

Conversely, the toroidal magnetic component undergoes continuous, robust amplification.
As the collapsing envelope forms a differentially rotating disk, the large-scale poloidal field lines are vigorously sheared in the azimuthal direction.
This magnetic winding (the $\Omega$-effect) efficiently uses the rotational kinetic energy of the disk, generating immense toroidal fields and driving the expansion of vertical magnetic towers, as illustrated by the $B^\phi$ spatial profiles and poloidal streamlines in Figure~\ref{fig:mag_Bphi_time_series}.

Analytically, the induction equation dictates that the linear growth of the toroidal field strength scales as $B_{\rm tor} \propto t B_{\rm pol} \partial_i \Omega$.
In the early, approximately kinematic winding phase, the integrated toroidal magnetic energy is therefore expected to scale approximately quadratically with time, $E_{B_{\rm tor}} \propto t^2$, as observed in our models.
Crucially, within the actively winding inner region of the disk, the spatial gradients of the angular velocity $(\partial_i \Omega)$ are similar for the two envelope rotation parameters considered here, $\eta=0.3$ and $0.5$.
Although the disks are geometrically thick and strongly pressure-supported, and their global density, pressure, mass, and angular-momentum distributions depend on $\eta$, this does not preclude similar local shear profiles in the principal winding region.
Thus, the similar early-time $t^2$ evolution of $E_{B_{\rm tor}}$ at fixed $B_{\rm surf}$ is consistent with the similar local shear measured in this region, rather than implying that the global disk structure or the extent of the magnetically affected region is independent of $\eta$.
Ultimately, because macroscopic magnetic winding is fully captured by our grid resolution, it operates as the dominant MHD mechanism in our simulations, driving the toroidal magnetic energy to values orders of magnitude greater than its poloidal counterpart.

A low-density, highly magnetised polar funnel—driven by sufficiently rapid rotation and strong magnetic fields—is an essential prerequisite for launching relativistic jets.
Figure~\ref{fig:mag_beta_mag_time_series} illustrates the inverse plasma beta ($\beta^{-1}_{\rm mag} = p_{\rm mag}/p_{\rm gas}$) profiles for the $\eta=0.3$ models at various times.
The $B_{\rm surf}= 4.64 \times 10^{10}$~G model is not shown, as its dynamics are entirely fluid-dominated ($\beta^{-1}_{\rm mag} \leq 10^{-4}$).
In the $B_{\rm surf}=4.64 \times 10^{12}$~G model, while the magnetic pressure initially rivals the gas pressure ($\beta_{\rm mag}\sim 1$) in the post-shock region near the poles, the gas pressure eventually overwhelms the magnetic field almost everywhere at later times.
Notably, regions where $\beta_{\rm mag}\sim 1$ persist around the equatorial surface of the NS, providing additional magnetic support that suppresses equatorial accretion, but this magnetic dominance fails to establish itself along the polar axis.

In models with stronger initial magnetisation, a magnetically dominated funnel successfully forms along the poles following the initial transient phase.
For the $B_{\rm surf} = 2.32 \times 10^{13}$~G model, this jet-like structure reaches peak magnetic dominance around $t \sim 80$~ms, though it weakens at later times.
Conversely, the $B_{\rm surf} = 4.64 \times 10^{13}$~G model sustains a much wider and more robust magnetically dominated funnel.

\begin{figure*}
    \centering
    \includegraphics[width=\textwidth, angle=0]{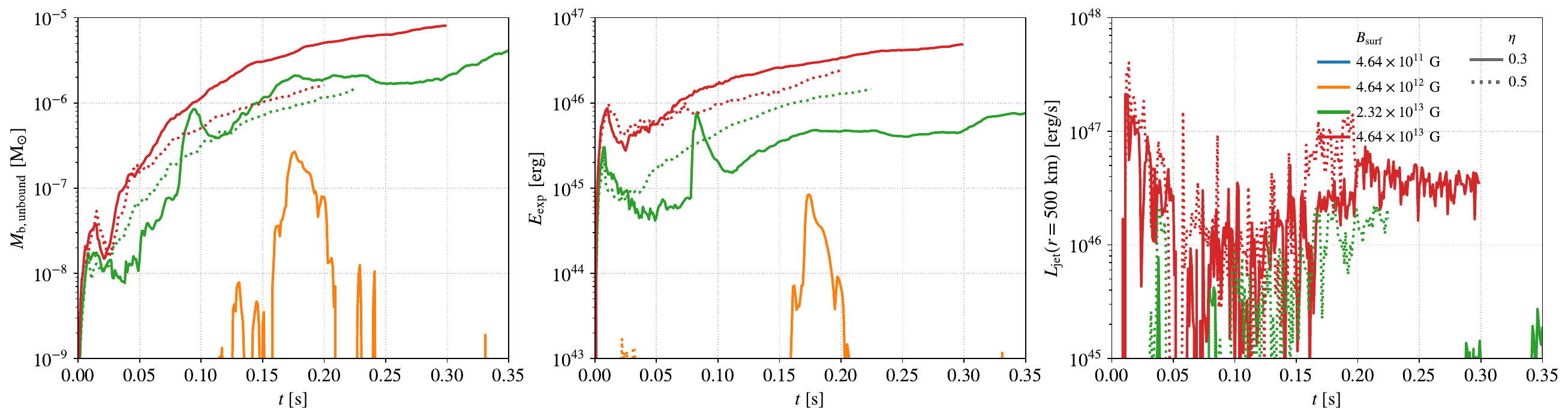}
    \caption{
	\emph{Left and middle panels}: Time evolution of the unbounded rest mass and the energy in the post-shock regions.
        \emph{Right panel}: Time evolution of the jet power extracted at $r=500$~km for all magnetised cases.
    }
    \label{fig:mag_ub_jet_vs_t}
\end{figure*}

Right panel of the Figure~\ref{fig:mag_ub_jet_vs_t} displays the temporal evolution of the jet power extracted at $r=500$~km for the high-$B$ models across all values of $\eta$.
Low-$B$ models are not shown from this analysis because they fail to form funnels with sufficient magnetisation ($\sigma_{\rm mag} > 1$) at the extraction radius to drive an outflow.
For the most strongly magnetised cases ($B_{\rm surf} = 4.64 \times 10^{13}$~G), regardless of $\eta$, a jet successfully breaks out past the extraction radius, sustaining a jet power on the order of $\mathcal{O}(10^{46})$~erg/s.

In contrast, the jets in the $B_{\rm surf} = 2.32 \times 10^{13}$ G models struggle to overcome the ram pressure of the infalling envelope, exhibiting `choked' or stuttering behaviours.
For both $\eta$ values, an initial short burst occurs at $t \sim 0.03$~s, lasting less than $0.02$~s before falling back.
At $t \sim 0.08$~s, a secondary burst arises.
In the $\eta=0.3$ case, this secondary burst is again extremely short-lived, quickly collapsing with no further breakouts (i.e. $L_{\rm jet} \gtrsim 2 \times 10^{45}~{\rm erg/s}$) recorded at this radius during our simulated time.
However, in the $\eta=0.5$ case, the jet successfully revives from this secondary burst, resulting in a more sustained outflow with a power slightly below $10^{46}$~erg/s.

\begin{figure*}
    \centering
    \includegraphics[width=\textwidth, angle=0]{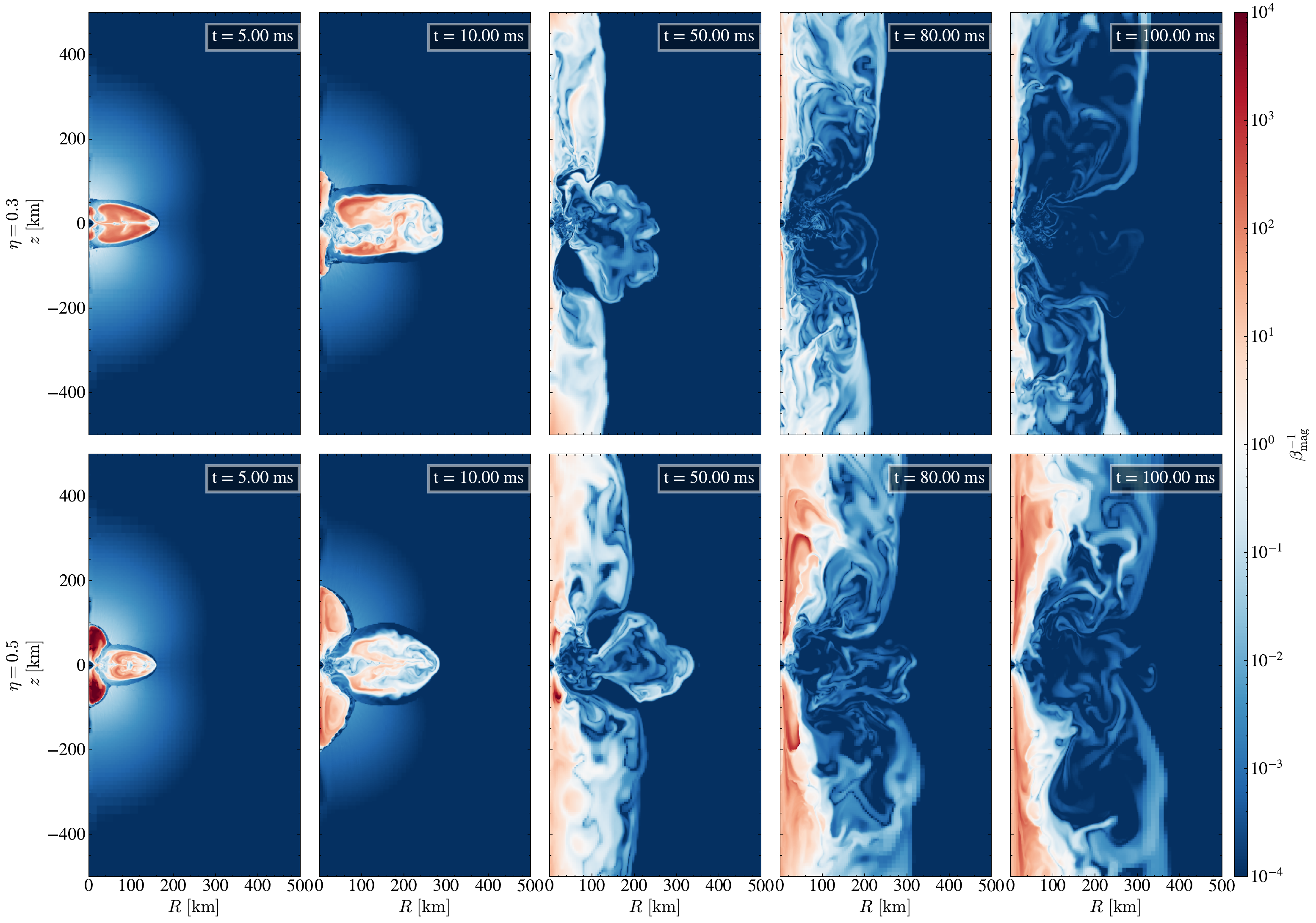}
    \caption{
	    Inverse plasma beta ($\beta^{-1}_{\rm mag}$) profiles comparing $B_{\rm surf}= 2.32 \times 10^{13}~{\rm G}$ models with different rotation models at various times, highlighting the effect of centrifugal funnel evacuation.
    }
    \label{fig:mag_beta_mag_time_series_compare_eta}
\end{figure*}

The divergent fates of the jets in the $B_{\rm surf} = 2.32 \times 10^{13}$~G models highlight the critical role of the envelope's rotation.
A faster-rotating envelope ($\eta=0.5$) carries greater specific angular momentum, subjecting the infalling gas to a stronger centrifugal barrier (as detailed for the non-magnetised cases in Section~\ref{sec:rot}).
This enhanced centrifugal force naturally evacuates gas from the rotational axis, resulting in a polar funnel that is less baryon-loaded.
As compared in Figure~\ref{fig:mag_beta_mag_time_series_compare_eta}, the $\eta=0.5$ case exhibits a stronger initial burst at the poles and carves out a substantially wider magnetically dominated funnel than its $\eta=0.3$ counterpart.
By mitigating the ram pressure of the infalling baryons, the rapid rotation provides the ideal environmental conditions necessary for the jet to break out and survive at late times.

To evaluate the potential for a coincident optical transient, we estimate the total unbound mass and diagnostic explosion energy within the shocked region (see the left and middle panels of Figure~\ref{fig:mag_ub_jet_vs_t}).
Sustained growth of the unbound mass and energy is only achieved when the initial magnetic field is sufficiently strong ($B_{\rm surf} \geq 2.32 \times 10^{13}$~G).
While the $B_{\rm surf} = 4.64 \times 10^{12}$~G model exhibits a brief period where a fraction of the mass is unbound, this material is eventually recaptured and becomes gravitationally bound again at later times due to the continuously intensifying accretion flow.
In our most strongly magnetised cases, the unbound mass and energy saturate at approximately $\mathcal{O}(10^{-6})~M_\odot$ and $\mathcal{O}(10^{46})$~erg, respectively.
We note that these quantities are calculated instantaneously and locally within the post-shock region; their late-time evolution may deviate from these values, particularly as the macroscopic mass accretion continues to strengthen.
By the end of our simulations at $t \sim 0.35$ s, the forward shock of the outflow has propagated to a radius of approximately $r \sim 2,000$ km, see Figure~\ref{fig:mag_mdot_vs_t}.
Because this is still deep within the stellar interior (well below typical stripped-envelope radii of $\sim 10^5$ km), the outflow properties derived here strictly represent the early-time, engine-driven injection phase prior to jet breakout.

\begin{figure*}
    \centering
    \includegraphics[width=\textwidth, angle=0]{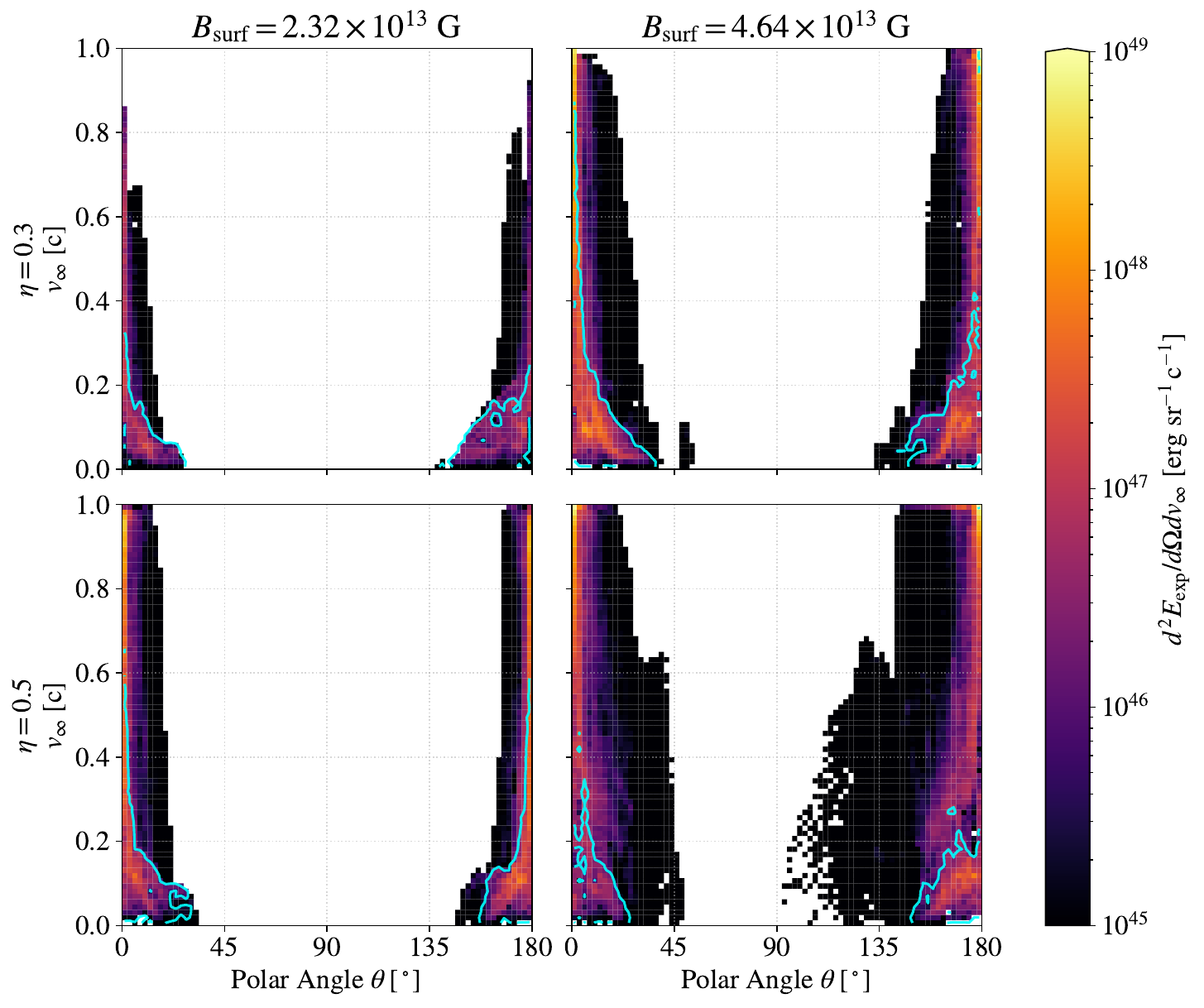}
    \caption{
	Two-dimensional phase space distributions of the differential diagnostic explosion energy (heatmap) and unbound ejecta mass (contours) as a function of polar angle $\theta$ and asymptotic velocity $v_\infty$. 
	The snapshots are taken at $t = 200$~ms for four different central engine models. 
	The columns compare different initial surface magnetic field strengths ($B_{\rm surf} = 2.32 \times 10^{13}$~G, \emph{left}; $4.64 \times 10^{13}$~G, \emph{right}), while the rows compare different rotation parameters ($\eta = 0.3$, \emph{top}; $\eta = 0.5$, \emph{bottom}). 
	The color map indicates the differential explosion energy density $d^2E_{\rm exp}/d\Omega dv_\infty$ in units of $\text{erg~sr}^{-1}\,c^{-1}$. 
	The solid cyan contours denote the differential mass distribution $d^2M_{\rm eje}/d\Omega dv_\infty$, highlighting the $10^{-6}~M_\odot~\text{sr}^{-1}\,c^{-1}$ level. 
	The figure clearly demonstrates a broad outflow structure: a massive, sub-relativistic wind/cocoon concentrated at low velocities ($v_\infty \lesssim 0.2c$) that gradually transitions into a highly energetic, low-mass, mildly relativistic polar jet, whose maximum velocity and total energy scale strongly with the magnetisation of the central engine.
    }
    \label{fig:mag_outflow_phase}
\end{figure*}

The velocity and angular distributions of the outflow are captured in Figure~\ref{fig:mag_outflow_phase}.
Overall, the outflow exhibits a clear bipolar structure, with almost no unbound matter found along the equatorial plane.
As shown in the figure, the vast majority of the unbound mass resides at low velocities ($v_\infty \lesssim 0.2c$), which is expected to later expand as a massive, sub-relativistic wind or cocoon.
In the cases where a jet is successfully launched, a highly energetic polar component is present despite its extremely low rest-mass density.
These polar jets are mildly to highly relativistic (reaching $v_\infty \gtrsim 0.8c$) and remain tightly collimated along a narrow opening angle.

\begin{figure*}
    \centering
    \includegraphics[width=\textwidth]{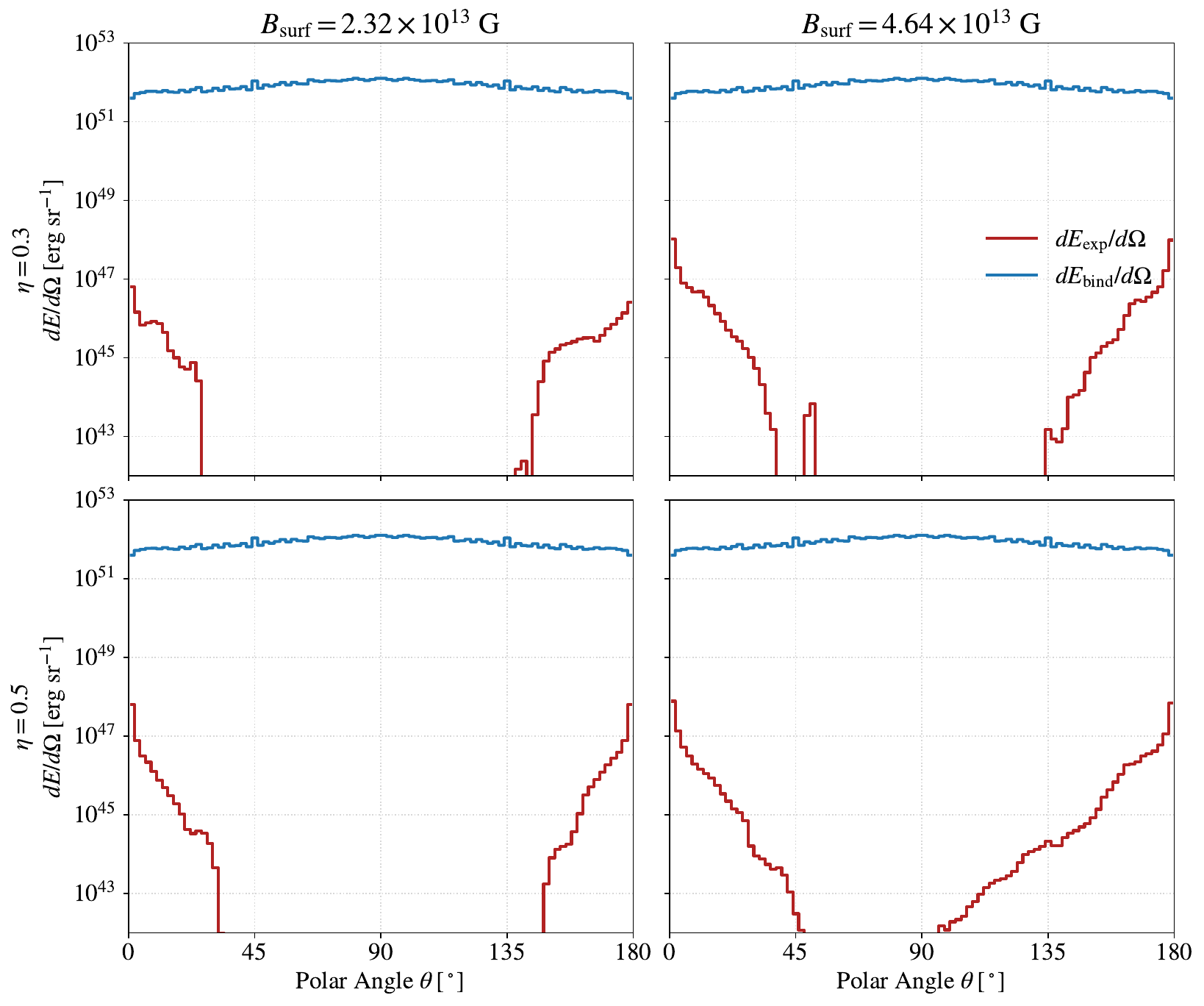}
    \caption{
	Angular density profiles of the diagnostic explosion energy {$dE_{\rm exp}/d\Omega$} (red step curves) compared to the gravitational binding energy density {$dE_{\rm bind}/d\Omega$} of the overlying pre-shock stellar envelope (blue step curves) evaluated at {$t = 200$~ms}.
	Rows compare envelope rotation rates ($\eta = 0.3$, \emph{top}; $\eta = 0.5$, \emph{bottom}), while columns compare initial surface magnetic field strengths ($B_{\rm surf} = 2.32 \times 10^{13}$~G, \emph{left}; $4.64 \times 10^{13}$~G, \emph{right}).
	{The envelope binding energy density is nearly isotropic ($dE_{\rm bind}/d\Omega \approx 10^{52}~\text{erg}~\text{sr}^{-1}$), whereas energy injection is strongly concentrated toward the poles ($\theta \lesssim 30^\circ$), where $dE_{\rm exp}/d\Omega$ reaches $10^{46}$--$10^{47}~\text{erg}~\text{sr}^{-1}$.}
	Sustained engine activity beyond {$t = 200$~ms}, combined with common-envelope stripping in binary progenitors, suggests that the polar jet could break out while the equatorial/wide-angle wind remains choked inside the progenitor.
    }
    \label{fig:mag_outflow_1d_energy}
\end{figure*}

To assess whether the engine-driven outflows can successfully escape the progenitor star, Figure~\ref{fig:mag_outflow_1d_energy} compares the angular density of the diagnostic explosion energy $dE_{\rm exp}/d\Omega$ with the binding energy density $dE_{\rm bind}/d\Omega$ of the pre-shock stellar envelope.

While a global, spherical unbinding of the $15~M_\odot$ stellar envelope would require an integrated energy budget of $E_{\rm bind, total} \sim 10^{51}$~erg, a collimated polar outflow does not need to lift the entire star.
Instead, it only needs to overcome the gravitational overburden sitting directly within its narrow polar opening cone ($\theta \lesssim 10^\circ$).
In terms of angular distribution, the pre-shock envelope binding energy density is roughly isotropic at $dE_{\rm bind}/d\Omega \approx 10^{52}~\text{erg}~\text{sr}^{-1}$, whereas the engine deposits energy almost exclusively along the rotational axis, reaching $dE_{\rm exp}/d\Omega \sim 10^{46}$--$10^{47}~\text{erg}~\text{sr}^{-1}$ at $t = 200$~ms.

It is important to emphasize that the profiles in Figure~\ref{fig:mag_outflow_1d_energy} are evaluated at {$t = 200$~ms}, when the central engine has only just begun operating.
As hypercritical accretion continues over long physical timescales (${\sim} 1$--$2$~s), continuous energy injection could build up cumulative integrated energy at the shock front.
Therefore, while the wide-angle wind ($\theta \gtrsim 30^\circ$) likely remains choked within the dense envelope, the highly energetic polar jet is very likely to drill through the polar overburden and achieve successful shock breakout.

Furthermore, the unperturbed 1D stellar progenitor structure used in our setup represents a strict upper bound on the stellar overburden.
In a realistic CEE, orbital energy dissipation during the inspiral phase ejects a major fraction ($50\%$--$90\%$) of the envelope mass prior to central engine ignition.
This pre-ejection reduces the remaining envelope binding energy down to $E_{\rm bind} \sim 10^{47}$--$10^{49}$~erg, drastically lowering the physical threshold for successful jet breakout.

\begin{figure*}
    \centering
    \includegraphics[width=\textwidth, angle=0]{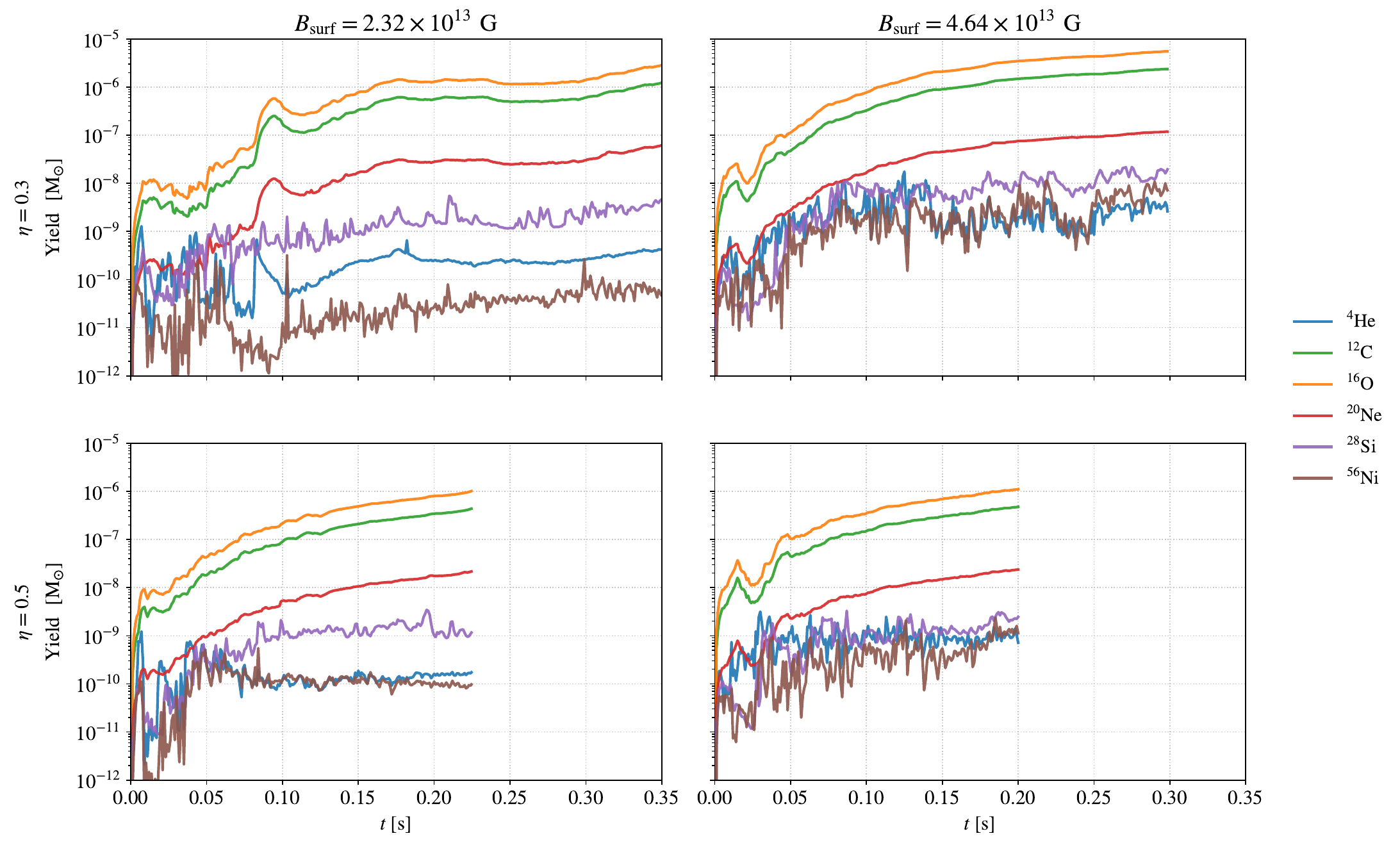}
    \caption{
    Time evolution of the nucleosynthetic yields for the gravitationally unbound ejecta across four central engine models. 
    The panels are arranged by the initial surface magnetic field strength ($B_{\rm surf} = 2.32 \times 10^{13}$ G, \emph{left column}; $4.64 \times 10^{13}$ G, \emph{right column}) and the engine rotation parameter ($\eta = 0.3$, \emph{top row}; $\eta = 0.5$, \emph{bottom row}). 
    The solid lines track the cumulative mass (in $M_\odot$) of selected isotopes: $^{4}$He, $^{12}$C, $^{16}$O, $^{20}$Ne, $^{28}$Si, and $^{56}$Ni. 
    Across all simulated parameter spaces, the composition of the unbound outflow is overwhelmingly dominated by unburned material from the original progenitor envelope (primarily $^{12}$C and $^{16}$O). 
    Conversely, the yields of intermediate-mass and iron-peak elements synthesised via explosive burning (e.g., $^{28}$Si and $^{56}$Ni) remain orders of magnitude lower. 
    This indicates that the early mass ejection is predominantly driven by the kinetic, magnetohydrodynamic sweep-up of the stellar envelope into a cocoon, rather than by a thermally-driven explosive shock.
    }
    \label{fig:mag_yields_vs_t}
\end{figure*}

Figure~\ref{fig:mag_yields_vs_t} compares the time evolution of the nucleosynthetic yields for the high-$B_{\rm surf}$ models.
The isotopes $^{12}$C and $^{16}$O strictly dominate the composition of the unbound material.
Note that this directly reflects the dominant initial composition of the stellar envelope, as discussed in \cite{2026PhRvD.114b3018C}.
While the shocked region is substantially hotter and denser than the unshocked envelope, a massive amount of pristine stellar material is rapidly swept up and accreted into this region at a rate of $\dot{M} \sim 10^{-3}~{\rm M_{\odot}/s}$, see Figure~\ref{fig:mag_mdot_vs_t}.
Crucially, the dynamical expansion timescale of this magnetohydrodynamically driven outflow is extremely short compared to the nuclear burning timescale.
Consequently, the outflow is mechanically accelerated before explosive nucleosynthesis can take place, leaving the composition largely unchanged from the progenitor profile.

Furthermore, the electron fraction $Y_{\rm e}$ of the ejecta remains firmly around $0.5$.
This behaves distinctly differently from cases involving the formation of a strongly magnetised proto-neutron star (PNS; e.g., \cite{2025ApJ...978L..38C}).
In our models, the central NS is uniformly rotating, which limits the efficacy of internal MHD processes compared to standard core-collapse supernovae or binary neutron star mergers, where the engine is highly differentially rotating.
Consequently, there is limited angular momentum transport acting to dredge up neutron-rich material from deep within the NS.
Because the stellar envelope natively begins with $Y_e = 0.5$, and the combination of mass accretion and neutrino irradiation is not perfectly tuned to drive massive neutronization within the polar funnel, the ejected outflow simply retains the pristine electron fraction of the progenitor star.

\section{Discussion}\label{sec:discussion}

\subsection{Accretion Dynamics and the Magnetic Barrier}
In this work, we present energy-integrated, two-moment neutrino transport GRMHD simulations of rotating, magnetised, hypercritical accretion onto a neutron star (NS) embedded within a massive stellar envelope.
We explore variations in the NS spin, initial magnetic field strength, and the rotation profile of the infalling envelope.
For the magnetisation, we consider initial polar field strengths of $B_{\rm surf} \in \{4.64 \times 10^{10}, 4.64 \times 10^{12}, 2.32 \times 10^{13}, 4.64 \times 10^{13}\}$~G at the pole of the NS.

From a dynamical perspective, the rotation of the envelope and the magnetic fields provide vital support against the infalling material.
Driven by the differentially rotating envelope and the initial poloidal magnetic fields, all magnetised models experience sustained toroidal magnetic field amplification.
This continuous winding forms a ``magnetic tower'' structure, which provides the necessary magnetic {stresses} to launch outflows.

Crucially, our simulations reveal that strong magnetisation fundamentally alters the boundary conditions of the central engine.
For models with $B_{\rm surf} \ge 2.32 \times 10^{13}$~G, the accumulated magnetic pressure, aided by the centrifugal barrier, becomes so intense that accretion onto the NS surface is completely halted.
Mass flowing inward from the envelope is temporarily trapped in a highly magnetised, turbulent reservoir just outside the magnetosphere, functionally decoupling the NS from the immediate infall and delaying its eventual collapse.

\subsection{Jet Launching and the Primordial Magnetic Field Limit}
In our most highly magnetised configurations ($B_{\rm surf} = 4.64 \times 10^{13}$~G), a structured, magnetically driven polar outflow---an incipient jet---is successfully launched, accompanied by unbound material.
The measured jet power is $\mathcal{O}(10^{46})$~erg/s.
While this power falls short of the ${\sim} 10^{50}$~erg/s required for a canonical long Gamma-Ray Burst (GRB), it represents a viable mechanism for powering a low-luminosity, engine-driven transient.

From an astronomical and evolutionary perspective, it is crucial to emphasise that a true primordial magnetar-level field ($B_{\rm surf} \gtrsim 10^{14}$~G) is not a prerequisite for launching these precursor outflows.
By the time a binary companion evolves into a red supergiant and engulfs the NS to initiate the common envelope phase, the NS is typically a canonical, ordinary pulsar with a surface magnetic field of $B_{\rm surf} \lesssim 10^{12}$~G.
Our simulation results demonstrate that for a successful jet breakout, the surface field only needs to reach $B_{\rm surf} \sim 2.32 \times 10^{13}$~G, provided the infalling envelope is rotating rapidly (e.g., our $\eta = 0.5$ models).
Transitioning from a standard pulsar field ($B_{\rm surf} \sim 10^{12}$~G) to this jet-launching threshold requires an amplification factor of only ${\sim} 10$.
Such a modest increase can be easily and rapidly achieved via the $\alpha$-$\Omega$ dynamo or continuous magnetic winding, fueled by the immense reservoir of differential rotation brought inward by the common envelope itself.

Conversely, exploratory test runs initialised with extreme primordial fields (e.g., $B_{\rm surf} \gtrsim 5 \times 10^{14}$~G) were unable to establish a stable accretion state.
In these cases, the overwhelming magnetic pressure violently disrupted the initial equilibrium of the inner envelope, leading to severe numerical instabilities before a steady flow could form.
This physical and numerical constraint strongly implies that the transient engine relies on a completely ordinary pulsar that dynamically scales up its field to a critical threshold via the accretion flow, rather than a pre-existing magnetar remnant.

\subsection{Physical Outcomes of the Central Engine}\label{sec:discussion_outcomes}

A central takeaway from our simulations is that the diagnostic explosion energy injected by the strongest magnetised NS engine ($E_{\rm exp} \sim \mathcal{O}(10^{45})$~erg) and the associated unbound mass ($\sim \mathcal{O}(10^{-6})~M_\odot$) are fundamentally insufficient to unbind the massive envelope of a red supergiant.
Therefore, rather than driving the primary supernova, this hyperaccreting NS phase must be viewed strictly as a transient precursor stage.
Because the system is fueled by an effectively infinite mass reservoir, the NS will eventually exceed its maximum mass limit and collapse into a BH.
The ultimate observational signature of the transient depends on how this NS precursor phase modifies the stellar interior before BH formation.

Based on the outflow dynamics observed in our data, we anticipate three evolutionary pathways:

\paragraph{Weakly Magnetised Engine: Quiet Collapse} 
In models with weak initial magnetic fields (e.g., $B_{\rm surf} \le 4.64 \times 10^{12}$~G) or insufficient rotation, the central engine fails to launch an outflow or establish a magnetic barrier.
The mass falls freely onto the NS, which is rapidly overwhelmed by the accretion flow.
In this scenario, the system undergoes a prompt, quiet collapse into a BH \cite{2026PhRvD.114b3018C}.
Observationally, this manifests as a ``failed supernova''; surveys would observe the sudden disappearance of a red supergiant, accompanied only by a thermal neutrino burst and a faint, long-wavelength transient as the envelope is consumed.
While the initial collapse itself remains quiet due to weak early field strengths, late-time accretion onto the newly formed black hole provides a secondary channel for activity.
Continued fallback accretion can progressively amplify magnetic fields in the inner disk via shear and flux accumulation, potentially launching delayed, late-time outflows long after the prompt NS phase has ended.

\paragraph{Moderately Magnetised Engine: Choked Jets} 
In moderately magnetised models with insufficient rotation (e.g., $B_{\rm surf} = 2.32 \times 10^{13}$~G and $\eta=0.3$), the magnetic field is strong enough to halt accretion at the surface, but the resulting jet struggles to overcome the ram pressure of the infalling envelope and ultimately stutters or chokes.
The ${\sim} 10^{45}$~erg of energy does not break out; instead, it is dumped into the stellar core, inflating a hot, pressurised cocoon.
Because the accretion is temporarily halted, the NS survives longer in a suspended state.
When the magnetic barrier eventually breaks and the NS collapses, the newly formed BH engine inherits this lower-density, high-entropy cocoon environment, which {may} drastically alter the dynamics of the subsequent primary explosion.
{Since our simulations do not track the post-collapse phase, future work is required to determine how this pre-cleared cavity might alter the dynamics of a subsequent explosion.}

\paragraph{Highly Magnetised Engine: Potential Polar Breakouts} 
In our most optimal models ($B_{\rm surf} = 4.64 \times 10^{13}$~G), the accretion is successfully halted, and a stable $\mathcal{O}(10^{46})$~erg/s jet successfully breaks out of the extraction region.
Observationally, because the energy budget is too low to unbind the star, this will not present as a canonical supernova.
Instead, observers aligned with the jet axis would detect a low-luminosity GRB (llGRB) or an extended X-ray precursor.
Crucially, this weak jet acts as a hydrodynamic drill, clearing a low-density polar funnel through the massive envelope.
When the NS subsequently collapses, the vastly more powerful BH-disk engine {could} launch its outflow directly into this pre-cleared channel, providing a highly efficient, virtually drag-free pathway for powering a classical long GRB.

These strongly magnetised models produce a fast, low-mass outflow propagating at $\sim 0.1~c$ along the polar axis, phenomenologically resembling the polar CEJSN scenario proposed by \cite{2019MNRAS.484.4972S}.
However, due to this sub-relativistic jet head velocity, a successful breakout within the engine's lifetime strictly requires a compact envelope.
Such conditions could be met if the neutron star merges with a heavily stripped massive star, or if a significant fraction of the extended envelope was already unbound by orbital drag during the prior inspiral phase.

\subsection{Observational Implications of a Choked Engine}\label{sec:discussion_observations}
As detailed in Section~\ref{sec:discussion_outcomes}, our simulations indicate that the jet power generated by the central engine ($\sim 10^{46}$ erg/s) is generally overwhelmed by the massive common envelope, making a choked jet the most likely physical outcome.
Therefore, we focus our synthetic observable predictions on this dominant channel.
Even if the outflow is insufficient to completely unbind the star, it is often energetic enough to drive a shock through the envelope, contributing additional expansion energy.
The orbital energy loss from the binary inspiral already unbinds much of the stellar envelope, and this expansion can power a transient.
There are many studies matching this base emission to luminous red novae \citep{2017ApJ...834..107B,2021MNRAS.508.2386S,2025ApJ...982...83H,2025ApJ...994L..41K,2026arXiv260626495M,2026arXiv260715390S}.
Here, we study the additional observational effect that the choked jets produced in our models have on these systems.

The choked explosion sends a shock through the star and, when it breaks out of the photosphere, this shock-heated material can produce blackbody emission in the UV and X-rays.
To estimate this emission, we use the simple model developed by \cite{2026arXiv260300820F} that parameterizes the properties of the ejecta by employing a distribution of ejecta velocities (which, in turn, yield the emitting temperature) and emitting areas for material moving at different velocities.
The distribution is described with a power law ($\propto \Gamma^{-p}$) with a peak Lorentz factor (for our models, we assume a weak shock and the peak Lorentz factor is 2).
The total energy, $E$, is the kinetic energy of the ejecta moving above $0.1\,c$ (where $c$ is the speed of light).
Although the emitting area can vary for different ejecta velocities, we assume this area, $A(\Gamma)$, is a constant ($A$).
{The injected ejecta mass and energy are guided by the results of our 2D GRMHD simulations. 
To account for theoretical uncertainties and bracket the expected light curves, we allow these quantities to vary around their simulation-guided baseline values.}

\begin{figure}
    \centering
    \includegraphics[width=\columnwidth]{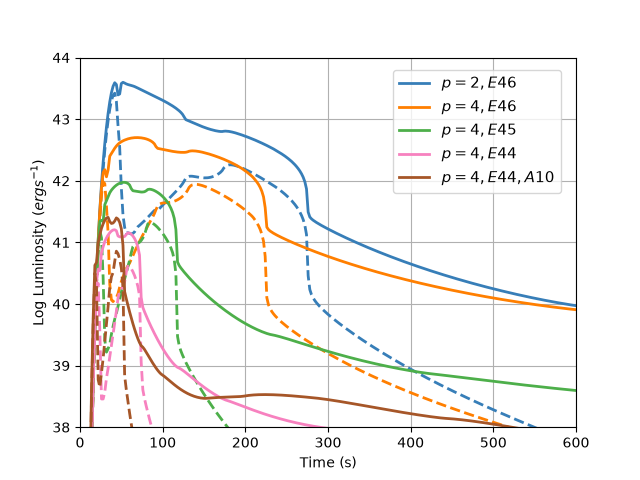}
    \caption{
	X-ray luminosity as a function of time above 0.1\,keV (solid) and above 0.3\,keV for a range of total shock energy ($10^{44}-10^{46} \, {\rm erg}$) and distribution of Lorentz factors $p=2-4$ where the peak Lorentz factor is 2.
	For most of the models, we assume a constant across Lorentz factor emitting area of $4\times10^{23} \, {\rm cm^2}$.
	We include one model where the emitting area is 10 times higher (A10).
	}
    \label{fig:sbo}
\end{figure}

Figure~\ref{fig:sbo} shows the X-ray luminosity for models we believe approximate our explosion.
With the very uncertain structure of the common envelope expansion, none of these parameters are well known.
We assume a fairly steep power-law distribution (combined with a peak Lorentz factor of 2, our model assumes very little fast ejecta).
Given the rapid expansion and asymmetric explosion, it is also difficult to estimate the emitting area.
As such, we vary these values considerably to determine their effect.
Except for our most energetic case, the peak luminosity (especially above 0.3\,keV) is fairly low.
Nonetheless, such X-ray flares could be detected by current detectors, although it is likely that the rate is 100 times lower than normal supernovae (Abrahams et al., in prep).

Although much less frequent than X-ray flares from supernovae, the broad-wavelength emission from these mergers will be very different than supernovae.
As discussed above, the ejecta from the common envelope inspiral is expected to produce a luminous red nova.
Our accretion-disk-driven explosion can add additional power to the transient (both by producing radioactive $^{56}$Ni and through shock interactions).
However, our models produce very little $^{56}$Ni ($< 10^{-9} \, M_\odot$), and the low energy of the ejecta produces weak shocks.

\begin{figure}
    \centering
    \includegraphics[width=\columnwidth]{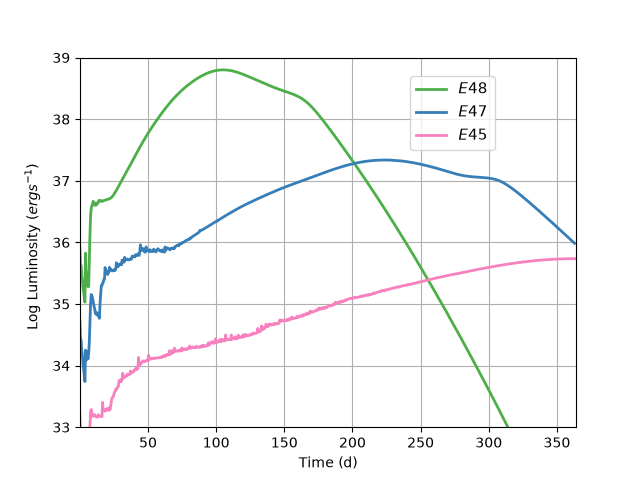}
    \caption{
	Luminosity versus time modeling the light curve produced by our choked jet models varying the amount of energy in the explosion ejecta.
	There is very little $^{56}$Ni in this ejecta, so the dominant power source is shock interactions.
	We assume shock interactions between our choked jet explosion and the expanding common  envelope convert 30\% of the kinetic energy into thermal energy over the first 10\,d using the prescription by \cite{2025ApJ...994..259N}.
	The slow moving ejecta produces extended but dim light curves.
	}
    \label{fig:fbollc}
\end{figure}

Figure~\ref{fig:fbollc} shows the additional emission caused by this explosion using the {\tt SNLC} code, which includes a recipe for heating from shock interactions~\citep{2025ApJ...994..259N}.
The additional optical/IR emission from this explosion is minimal.
Ultimately, these choked engine models will produce X-ray flares without bright optical or IR emission beyond the baseline luminous red nova.
We note that this specific scenario occurs in compact object common envelope systems and should account for only a small fraction of all observed luminous red novae.

\subsection{Limitations and Future Work}

Ultimately, our results highlight the difficulty of powering the most energetic CEJSN directly from a hyperaccreting NS \cite{2019MNRAS.484.4972S, 2025RAA....25b5023S}.
Within our 2D GRMHD framework, the crushing accretion rates and intense neutrino cooling overwhelmingly favour choked jets capped at $L_j \sim 10^{46} \text{ erg/s}$.
This severe energy deficit presents two distinct possibilities for the field.
On one hand, it suggests that the initial NS phase may serve merely as a transient precursor; generating the ${\sim} 10^{50} \text{ erg}$ outflows invoked by macroscopic models may strictly require the subsequent collapse into a BH, where a magnetically arrested disk can efficiently extract spin energy.
On the other hand, because 2D axisymmetry inherently suppresses the turbulent $\alpha$-dynamo, we cannot definitively rule out the NS engine.
It remains an open question whether fully 3D magnetic field amplification could generate sufficient magnetic pressure to halt the hypercritical accretion and launch a successful jet before collapse occurs.
The ultimate fate of these embedded central engines remains one of the most complex, unsolved problems in binary evolution, and bridging the gap between first-principles microphysics and macroscopic transients will require full 3D simulations to resolve.

As mentioned, a primary limitation of the current work is the assumption of 2D axisymmetry.
Axisymmetric simulations inherently suppress the $\alpha$-dynamo and limit the fully turbulent development of the MRI, thereby artificially restricting long-term magnetic field amplification.
To address this, future work will include incorporating phenomenological dynamo prescriptions (e.g., \cite{2025PhRvD.111l3017S}) or employing full 3D, high-resolution GRMHD configurations to better capture the complete $\alpha$-$\Omega$ dynamo and accurately track turbulent angular momentum transport.

Furthermore, we observe that strong magnetisation can temporarily decouple the infalling envelope from the NS surface, creating a stalled magnetospheric reservoir.
Predicting the exact timescale for the breakdown of this barrier, whether via magnetic interchange instabilities or direct gravitational crushing, is highly non-trivial.
Capturing the episodic drainage of this reservoir and determining the true survival time of the NS before BH collapse will require extended, seconds-long or even minutes-long 3D simulations.

Finally, the initial conditions governing the envelope rotation profile dictate the strength of the centrifugal barrier and the efficiency of the jet launching.
The idealised setups used in this study must eventually be replaced by initial conditions extracted directly from 3D CE evolution models, bridging the physical gap between the binary inspiral phase and the final core collapse.

\begin{acknowledgments}
DR acknowledges support from the Department of Energy, Office of Science, Division of Nuclear Physics under Awards Number DE-SC0024388, from the National Science Foundation under Grants No. PHY-2020275, PHY-2116686, PHY-2407681, PHY-2512802, and PHY-2621752.
The simulations in this work have been performed on Frontera at Texas Advanced Computing Center (TACC) under allocation no.~\#PHY23001. 

We have modified \texttt{RNS}~\citep{1995ApJ...444..306S} to generate the NSs.
The Modules for Experiments in Stellar Astrophysics \citep[MESA][]{Paxton2011, Paxton2013, Paxton2015, Paxton2018, Paxton2019, Jermyn2023} (version 25.12.1) is also used to generate massive star models.
The results of this work were produced by utilising the GRMHD code \texttt{Gmunu}~\citep{2020CQGra..37n5015C, 2021MNRAS.508.2279C}, with the grey M1 neutrino transport~\citep{2023ApJS..267...38C, 2024ApJ...975..116C} and the nuclear burning~\citep{2026ApJS..284....6C} modules.
The tabulated neutrino interactions were provided by \texttt{NuLib}~\citep{2015ApJS..219...24O}.
The EoS in the non-NSE region is provided by \texttt{helmeos}~\citep{2000ApJS..126..501T}. 
The nuclear networks used in this work (i.e. the \texttt{aprox13})~\citep{1999ApJS..124..241T, 2000ApJS..129..377T} are publicly available at \url{https://cococubed.com/}.
The data of the simulations were post-processed and visualised with 
\texttt{yt}~\citep{2011ApJS..192....9T},
\texttt{NumPy}~\citep{harris2020array}, 
\texttt{pandas}~\citep{reback2020pandas, mckinney-proc-scipy-2010},
\texttt{SciPy}~\citep{2020SciPy-NMeth} and
\texttt{Matplotlib}~\citep{2007CSE.....9...90H, thomas_a_caswell_2023_7697899}.
\end{acknowledgments}

\appendix

% \nocite{*}

% \bibliographystyle{apsrev4-2}
% \bibliographystyle{aas_marcos}
\bibliography{references}{}% Produces the bibliography via BibTeX.

@ARTICLE{2020CQGra..37n5015C,
       author = {{Cheong}, Patrick Chi-Kit and {Lin}, Lap-Ming and {Li}, Tjonnie Guang Feng},
        title = "{Gmunu: toward multigrid based Einstein field equations solver for general-relativistic hydrodynamics simulations}",
      journal = {Classical and Quantum Gravity},
         year = 2020,
        month = jul,
       volume = {37},
       number = {14},
          eid = {145015},
        pages = {145015},
          doi = {10.1088/1361-6382/ab8e9c},
archivePrefix = {arXiv},
       eprint = {2001.05723},
 primaryClass = {gr-qc},
       adsurl = {https://ui.adsabs.harvard.edu/abs/2020CQGra..37n5015C}
}

@ARTICLE{2021MNRAS.508.2279C,
       author = {{Cheong}, Patrick Chi-Kit and {Lam}, Alan Tsz-Lok and {Ng}, Harry Ho-Yin and {Li}, Tjonnie Guang Feng},
        title = "{Gmunu: paralleled, grid-adaptive, general-relativistic magnetohydrodynamics in curvilinear geometries in dynamical space-times}",
      journal = {\mnras},
         year = 2021,
        month = dec,
       volume = {508},
       number = {2},
        pages = {2279-2301},
          doi = {10.1093/mnras/stab2606},
archivePrefix = {arXiv},
       eprint = {2012.07322},
 primaryClass = {astro-ph.IM},
       adsurl = {https://ui.adsabs.harvard.edu/abs/2021MNRAS.508.2279C}
}

@ARTICLE{2022ApJS..261...22C,
       author = {{Cheong}, Patrick Chi-Kit and {Pong}, David Yat Tung and {Yip}, Anson Ka Long and {Li}, Tjonnie Guang Feng},
        title = "{An Extension of Gmunu: General-relativistic Resistive Magnetohydrodynamics Based on Staggered-meshed Constrained Transport with Elliptic Cleaning}",
      journal = {\apjs},
         year = 2022,
        month = aug,
       volume = {261},
       number = {2},
          eid = {22},
        pages = {22},
          doi = {10.3847/1538-4365/ac6cec},
archivePrefix = {arXiv},
       eprint = {2110.03732},
 primaryClass = {astro-ph.IM},
       adsurl = {https://ui.adsabs.harvard.edu/abs/2022ApJS..261...22C}
}

@ARTICLE{2023ApJS..267...38C,
       author = {{Cheong}, Patrick Chi-Kit and {Ng}, Harry Ho-Yin and {Lam}, Alan Tsz-Lok and {Li}, Tjonnie Guang Feng},
        title = "{General-relativistic Radiation Transport Scheme in Gmunu. I. Implementation of Two-moment-based Multifrequency Radiative Transfer and Code Tests}",
      journal = {\apjs},
         year = 2023,
        month = aug,
       volume = {267},
       number = {2},
          eid = {38},
        pages = {38},
          doi = {10.3847/1538-4365/acd931},
archivePrefix = {arXiv},
       eprint = {2303.03261},
 primaryClass = {astro-ph.IM},
       adsurl = {https://ui.adsabs.harvard.edu/abs/2023ApJS..267...38C}
}

@ARTICLE{2024ApJ...975..116C,
       author = {{Cheong}, Patrick Chi-Kit and {Foucart}, Francois and {Duez}, Matthew D. and {Offermans}, Arthur and {Muhammed}, Nishad and {Chawhan}, Pavan},
        title = "{Energy-dependent and Energy-integrated Two-moment General-relativistic Neutrino Transport Simulations of a Hypermassive Neutron Star}",
      journal = {\apj},
         year = 2024,
        month = nov,
       volume = {975},
       number = {1},
          eid = {116},
        pages = {116},
          doi = {10.3847/1538-4357/ad7825},
archivePrefix = {arXiv},
       eprint = {2407.16017},
 primaryClass = {astro-ph.HE},
       adsurl = {https://ui.adsabs.harvard.edu/abs/2024ApJ...975..116C}
}

@ARTICLE{2026ApJS..284....6C,
       author = {{Cheong}, Patrick Chi-Kit and {Fryer}, Christopher L.},
        title = "{Toward First-principles Multimessenger Predictions: Coupling Nuclear Networks with General Relativistic Radiation Magnetohydrodynamics in Gmunu}",
      journal = {\apjs},
         year = 2026,
        month = may,
       volume = {284},
       number = {1},
          eid = {6},
        pages = {6},
          doi = {10.3847/1538-4365/ae48e4},
archivePrefix = {arXiv},
       eprint = {2510.12978},
 primaryClass = {astro-ph.IM},
       adsurl = {https://ui.adsabs.harvard.edu/abs/2026ApJS..284....6C}
}

@ARTICLE{1995ApJ...444..306S,
       author = {{Stergioulas}, Nikolaos and {Friedman}, John L.},
        title = "{Comparing Models of Rapidly Rotating Relativistic Stars Constructed by Two Numerical Methods}",
      journal = {\apj},
         year = 1995,
        month = may,
       volume = {444},
        pages = {306},
          doi = {10.1086/175605},
archivePrefix = {arXiv},
       eprint = {astro-ph/9411032},
 primaryClass = {astro-ph},
       adsurl = {https://ui.adsabs.harvard.edu/abs/1995ApJ...444..306S}
}

@ARTICLE{2015ApJS..219...24O,
       author = {{O'Connor}, Evan},
        title = "{An Open-source Neutrino Radiation Hydrodynamics Code for Core-collapse Supernovae}",
      journal = {\apjs},
         year = 2015,
        month = aug,
       volume = {219},
       number = {2},
          eid = {24},
        pages = {24},
          doi = {10.1088/0067-0049/219/2/24},
archivePrefix = {arXiv},
       eprint = {1411.7058},
 primaryClass = {astro-ph.HE},
       adsurl = {https://ui.adsabs.harvard.edu/abs/2015ApJS..219...24O}
}

@ARTICLE{2000ApJS..126..501T,
       author = {{Timmes}, F.~X. and {Swesty}, F. Douglas},
        title = "{The Accuracy, Consistency, and Speed of an Electron-Positron Equation of State Based on Table Interpolation of the Helmholtz Free Energy}",
      journal = {\apjs},
         year = 2000,
        month = feb,
       volume = {126},
       number = {2},
        pages = {501-516},
          doi = {10.1086/313304},
       adsurl = {https://ui.adsabs.harvard.edu/abs/2000ApJS..126..501T}
}

@ARTICLE{1999ApJS..124..241T,
       author = {{Timmes}, F.~X.},
        title = "{Integration of Nuclear Reaction Networks for Stellar Hydrodynamics}",
      journal = {\apjs},
         year = 1999,
        month = sep,
       volume = {124},
       number = {1},
        pages = {241-263},
          doi = {10.1086/313257},
       adsurl = {https://ui.adsabs.harvard.edu/abs/1999ApJS..124..241T}
}

@ARTICLE{2000ApJS..129..377T,
       author = {{Timmes}, F.~X. and {Hoffman}, R.~D. and {Woosley}, S.~E.},
        title = "{An Inexpensive Nuclear Energy Generation Network for Stellar Hydrodynamics}",
      journal = {\apjs},
         year = 2000,
        month = jul,
       volume = {129},
       number = {1},
        pages = {377-398},
          doi = {10.1086/313407},
       adsurl = {https://ui.adsabs.harvard.edu/abs/2000ApJS..129..377T}
}

@ARTICLE{2007CSE.....9...90H,
       author = {{Hunter}, John D.},
        title = "{Matplotlib: A 2D Graphics Environment}",
      journal = {Computing in Science and Engineering},
         year = 2007,
        month = may,
       volume = {9},
       number = {3},
        pages = {90-95},
          doi = {10.1109/MCSE.2007.55},
       adsurl = {https://ui.adsabs.harvard.edu/abs/2007CSE.....9...90H}
}

@software{thomas_a_caswell_2023_7697899,
  author       = {Thomas A Caswell and
                  Antony Lee and
                  Elliott Sales de Andrade and
                  Michael Droettboom and
                  Tim Hoffmann and
                  Jody Klymak and
                  John Hunter and
                  Eric Firing and
                  David Stansby and
                  Nelle Varoquaux and
                  Jens Hedegaard Nielsen and
                  Benjamin Root and
                  Ryan May and
                  Oscar Gustafsson and
                  Phil Elson and
                  Jouni K. Seppänen and
                  Jae-Joon Lee and
                  Darren Dale and
                  hannah and
                  Damon McDougall and
                  Andrew Straw and
                  Paul Hobson and
                  Kyle Sunden and
                  Greg Lucas and
                  Christoph Gohlke and
                  Adrien F. Vincent and
                  Tony S Yu and
                  Eric Ma and
                  Steven Silvester and
                  Charlie Moad},
  title        = {matplotlib/matplotlib: REL: v3.7.1},
  month        = mar,
  year         = 2023,
  publisher    = {Zenodo},
  version      = {v3.7.1},
  doi          = {10.5281/zenodo.7697899},
  url          = {https://doi.org/10.5281/zenodo.7697899}
}

@Article{harris2020array,
 title         = {Array programming with {NumPy}},
 author        = {Charles R. Harris and K. Jarrod Millman and St{\'{e}}fan J.
                 van der Walt and Ralf Gommers and Pauli Virtanen and David
                 Cournapeau and Eric Wieser and Julian Taylor and Sebastian
                 Berg and Nathaniel J. Smith and Robert Kern and Matti Picus
                 and Stephan Hoyer and Marten H. van Kerkwijk and Matthew
                 Brett and Allan Haldane and Jaime Fern{\'{a}}ndez del
                 R{\'{i}}o and Mark Wiebe and Pearu Peterson and Pierre
                 G{\'{e}}rard-Marchant and Kevin Sheppard and Tyler Reddy and
                 Warren Weckesser and Hameer Abbasi and Christoph Gohlke and
                 Travis E. Oliphant},
 year          = {2020},
 month         = sep,
 journal       = {Nature},
 volume        = {585},
 number        = {7825},
 pages         = {357--362},
 doi           = {10.1038/s41586-020-2649-2},
 publisher     = {Springer Science and Business Media {LLC}},
 url           = {https://doi.org/10.1038/s41586-020-2649-2}
}

@software{reback2020pandas,
    author       = {The pandas development team},
    title        = {pandas-dev/pandas: Pandas},
    month        = feb,
    year         = 2020,
    publisher    = {Zenodo},
    version      = {latest},
    doi          = {10.5281/zenodo.3509134},
    url          = {https://doi.org/10.5281/zenodo.3509134}
}

@InProceedings{ mckinney-proc-scipy-2010,
  author    = { {W}es {M}c{K}inney },
  title     = { {D}ata {S}tructures for {S}tatistical {C}omputing in {P}ython },
  booktitle = { {P}roceedings of the 9th {P}ython in {S}cience {C}onference },
  pages     = { 56 - 61 },
  year      = { 2010 },
  editor    = { {S}t\'efan van der {W}alt and {J}arrod {M}illman },
  doi       = { 10.25080/Majora-92bf1922-00a }
}

@ARTICLE{2020SciPy-NMeth,
  author  = {Virtanen, Pauli and Gommers, Ralf and Oliphant, Travis E. and
            Haberland, Matt and Reddy, Tyler and Cournapeau, David and
            Burovski, Evgeni and Peterson, Pearu and Weckesser, Warren and
            Bright, Jonathan and {van der Walt}, St{\'e}fan J. and
            Brett, Matthew and Wilson, Joshua and Millman, K. Jarrod and
            Mayorov, Nikolay and Nelson, Andrew R. J. and Jones, Eric and
            Kern, Robert and Larson, Eric and Carey, C J and
            Polat, {\.I}lhan and Feng, Yu and Moore, Eric W. and
            {VanderPlas}, Jake and Laxalde, Denis and Perktold, Josef and
            Cimrman, Robert and Henriksen, Ian and Quintero, E. A. and
            Harris, Charles R. and Archibald, Anne M. and
            Ribeiro, Ant{\^o}nio H. and Pedregosa, Fabian and
            {van Mulbregt}, Paul and {SciPy 1.0 Contributors}},
  title   = {{{SciPy} 1.0: Fundamental Algorithms for Scientific
            Computing in Python}},
  journal = {Nature Methods},
  year    = {2020},
  volume  = {17},
  pages   = {261--272},
  adsurl  = {https://rdcu.be/b08Wh},
  doi     = {10.1038/s41592-019-0686-2},
}

@ARTICLE{2011ApJS..192....9T,
   author = {{Turk}, M.~J. and {Smith}, B.~D. and {Oishi}, J.~S. and {Skory}, S. and
     {Skillman}, S.~W. and {Abel}, T. and {Norman}, M.~L.},
    title = "{yt: A Multi-code Analysis Toolkit for Astrophysical Simulation Data}",
  journal = {The Astrophysical Journal Supplement Series},
archivePrefix = "arXiv",
   eprint = {1011.3514},
 primaryClass = "astro-ph.IM",
     year = 2011,
    month = jan,
   volume = 192,
      eid = {9},
    pages = {9},
      doi = {10.1088/0067-0049/192/1/9},
   adsurl = {https://ui.adsabs.harvard.edu/abs/2011ApJS..192....9T}
}

@ARTICLE{Paxton2011,
  author = {{Paxton}, B. and {Bildsten}, L. and {Dotter}, A. and {Herwig}, F. and {Lesaffre}, P. and {Timmes}, F.},
  title = {{Modules for Experiments in Stellar Astrophysics (MESA)}},
  journal = {\apjs},
  archivePrefix = {arXiv},
  eprint = {1009.1622},
  primaryClass = {astro-ph.SR},
  year = {2011},
  month = {jan},
  volume = {192},
  eid = {3},
  pages = {3},
  doi = {10.1088/0067-0049/192/1/3},
  adsurl = {https://ui.adsabs.harvard.edu/abs/2011ApJS..192....3P},
}

@ARTICLE{Paxton2013,
  author = {{Paxton}, B. and {Cantiello}, M. and {Arras}, P. and {Bildsten}, L. and {Brown}, E.~F. and {Dotter}, A. and {Mankovich}, C. and {Montgomery}, M.~H. and {Stello}, D. and {Timmes}, F.~X. and {Townsend}, R.},
  title = {{Modules for Experiments in Stellar Astrophysics (MESA): Planets, Oscillations, Rotation, and Massive Stars}},
  journal = {\apjs},
  archivePrefix = {arXiv},
  eprint = {1301.0319},
  primaryClass = {astro-ph.SR},
  year = {2013},
  month = {sep},
  volume = {208},
  eid = {4},
  pages = {4},
  doi = {10.1088/0067-0049/208/1/4},
  adsurl = {https://ui.adsabs.harvard.edu/abs/2013ApJS..208....4P},
}

@ARTICLE{Paxton2015,
  author = {{Paxton}, B. and {Marchant}, P. and {Schwab}, J. and {Bauer}, E.~B. and {Bildsten}, L. and {Cantiello}, M. and {Dessart}, L. and {Farmer}, R. and {Hu}, H. and {Langer}, N. and {Townsend}, R.~H.~D. and {Townsley}, D.~M. and {Timmes}, F.~X.},
  title = {{Modules for Experiments in Stellar Astrophysics (MESA): Binaries, Pulsations, and Explosions}},
  journal = {\apjs},
  archivePrefix = {arXiv},
  eprint = {1506.03146},
  primaryClass = {astro-ph.SR},
  year = {2015},
  month = {sep},
  volume = {220},
  eid = {15},
  pages = {15},
  doi = {10.1088/0067-0049/220/1/15},
  adsurl = {https://ui.adsabs.harvard.edu/abs/2015ApJS..220...15P},
}

@ARTICLE{Paxton2018,
  author = {{Paxton}, B. and {Schwab}, J. and {Bauer}, E.~B. and {Bildsten}, L. and {Blinnikov}, S. and {Duffell}, P. and {Farmer}, R. and {Goldberg}, J.~A. and {Marchant}, P. and {Sorokina}, E. and {Thoul}, A. and {Townsend}, R.~H.~D. and {Timmes}, F.~X.},
  title = {{Modules for Experiments in Stellar Astrophysics (MESA): Convective Boundaries, Element Diffusion, and Massive Star Explosions}},
  journal = {\apjs},
  archivePrefix = {arXiv},
  eprint = {1710.08424},
  primaryClass = {astro-ph.SR},
  year = {2018},
  month = {feb},
  volume = {234},
  eid = {34},
  pages = {34},
  doi = {10.3847/1538-4365/aaa5a8},
  adsurl = {https://ui.adsabs.harvard.edu/abs/2018ApJS..234...34P},
}

@ARTICLE{Paxton2019,
       author = {{Paxton}, Bill and {Smolec}, R. and {Schwab}, Josiah and {Gautschy}, A. and
         {Bildsten}, Lars and {Cantiello}, Matteo and {Dotter}, Aaron and
         {Farmer}, R. and {Goldberg}, Jared A. and {Jermyn}, Adam S. and
         {Kanbur}, S.~M. and {Marchant}, Pablo and {Thoul}, Anne and
         {Townsend}, Richard H.~D. and {Wolf}, William M. and {Zhang}, Michael and
         {Timmes}, F.~X.},
        title = "{Modules for Experiments in Stellar Astrophysics (MESA): Pulsating Variable Stars, Rotation, Convective Boundaries, and Energy Conservation}",
      journal = {\apjs},
         year = "2019",
        month = "Jul",
       volume = {243},
       number = {1},
          eid = {10},
        pages = {10},
          doi = {10.3847/1538-4365/ab2241},
archivePrefix = {arXiv},
       eprint = {1903.01426},
 primaryClass = {astro-ph.SR},
       adsurl = {https://ui.adsabs.harvard.edu/abs/2019ApJS..243...10P}
}

@ARTICLE{Jermyn2023,
       author = {{Jermyn}, Adam S. and {Bauer}, Evan B. and {Schwab}, Josiah and {Farmer}, R. and {Ball}, Warrick H. and {Bellinger}, Earl P. and {Dotter}, Aaron and {Joyce}, Meridith and {Marchant}, Pablo and {Mombarg}, Joey S.~G. and {Wolf}, William M. and {Sunny Wong}, Tin Long and {Cinquegrana}, Giulia C. and {Farrell}, Eoin and {Smolec}, R. and {Thoul}, Anne and {Cantiello}, Matteo and {Herwig}, Falk and {Toloza}, Odette and {Bildsten}, Lars and {Townsend}, Richard H.~D. and {Timmes}, F.~X.},
        title = "{Modules for Experiments in Stellar Astrophysics (MESA): Time-dependent Convection, Energy Conservation, Automatic Differentiation, and Infrastructure}",
      journal = {\apjs},
         year = 2023,
        month = mar,
       volume = {265},
       number = {1},
          eid = {15},
        pages = {15},
          doi = {10.3847/1538-4365/acae8d},
archivePrefix = {arXiv},
       eprint = {2208.03651},
 primaryClass = {astro-ph.SR},
       adsurl = {https://ui.adsabs.harvard.edu/abs/2023ApJS..265...15J}
}

@ARTICLE{2010NuPhA.837..210H,
       author = {{Hempel}, Matthias and {Schaffner-Bielich}, J{\"u}rgen},
        title = "{A statistical model for a complete supernova equation of state}",
      journal = {\nphysa},
         year = 2010,
        month = jun,
       volume = {837},
       number = {3-4},
        pages = {210-254},
          doi = {10.1016/j.nuclphysa.2010.02.010},
archivePrefix = {arXiv},
       eprint = {0911.4073},
 primaryClass = {nucl-th},
       adsurl = {https://ui.adsabs.harvard.edu/abs/2010NuPhA.837..210H}
}

@ARTICLE{2024PhRvD.110d3015C,
       author = {{Cheong}, Patrick Chi-Kit and {Muhammed}, Nishad and {Chawhan}, Pavan and {Duez}, Matthew D. and {Foucart}, Francois and {Kidder}, Lawrence E. and {Pfeiffer}, Harald P. and {Scheel}, Mark A.},
        title = "{High angular momentum hot differentially rotating equilibrium star evolutions in conformally flat spacetime}",
      journal = {\prd},
         year = 2024,
        month = aug,
       volume = {110},
       number = {4},
          eid = {043015},
        pages = {043015},
          doi = {10.1103/PhysRevD.110.043015},
archivePrefix = {arXiv},
       eprint = {2402.18529},
 primaryClass = {astro-ph.HE},
       adsurl = {https://ui.adsabs.harvard.edu/abs/2024PhRvD.110d3015C}
}

@ARTICLE{1996PhRvD..53.5533C,
       author = {{Cook}, Gregory B. and {Shapiro}, Stuart L. and {Teukolsky}, Saul A.},
        title = "{Testing a simplified version of Einstein's equations for numerical relativity}",
      journal = {\prd},
         year = 1996,
        month = may,
       volume = {53},
       number = {10},
        pages = {5533-5540},
          doi = {10.1103/PhysRevD.53.5533},
archivePrefix = {arXiv},
       eprint = {gr-qc/9512009},
 primaryClass = {gr-qc},
       adsurl = {https://ui.adsabs.harvard.edu/abs/1996PhRvD..53.5533C}
}

@ARTICLE{2014GReGr..46.1800I,
       author = {{Iosif}, Panagiotis and {Stergioulas}, Nikolaos},
        title = "{On the accuracy of the IWM-CFC approximation in differentially rotating relativistic stars}",
      journal = {General Relativity and Gravitation},
         year = 2014,
        month = oct,
       volume = {46},
       number = {10},
          eid = {1800},
        pages = {1800},
          doi = {10.1007/s10714-014-1800-5},
archivePrefix = {arXiv},
       eprint = {1406.7375},
 primaryClass = {gr-qc},
       adsurl = {https://ui.adsabs.harvard.edu/abs/2014GReGr..46.1800I}
}

@ARTICLE{2021MNRAS.503..850I,
       author = {{Iosif}, Panagiotis and {Stergioulas}, Nikolaos},
        title = "{Equilibrium sequences of differentially rotating stars with post-merger-like rotational profiles}",
      journal = {\mnras},
         year = 2021,
        month = may,
       volume = {503},
       number = {1},
        pages = {850-866},
          doi = {10.1093/mnras/stab392},
archivePrefix = {arXiv},
       eprint = {2011.10612},
 primaryClass = {gr-qc},
       adsurl = {https://ui.adsabs.harvard.edu/abs/2021MNRAS.503..850I}
}

@ARTICLE{2022MNRAS.510.2948I,
       author = {{Iosif}, Panagiotis and {Stergioulas}, Nikolaos},
        title = "{Models of binary neutron star remnants with tabulated equations of state}",
      journal = {\mnras},
         year = 2022,
        month = feb,
       volume = {510},
       number = {2},
        pages = {2948-2967},
          doi = {10.1093/mnras/stab3565},
archivePrefix = {arXiv},
       eprint = {2104.13672},
 primaryClass = {astro-ph.HE},
       adsurl = {https://ui.adsabs.harvard.edu/abs/2022MNRAS.510.2948I}
}

@ARTICLE{2026PhRvD.114b3018C,
       author = {{Cheong}, Patrick Chi-Kit and {Radice}, David and {Fryer}, Christopher L.},
        title = "{Hyperaccreting neutron stars inside massive envelopes: The implausibility of Thorne-{\.Z}ytkow objects}",
      journal = {\prd},
         year = 2026,
        month = jul,
       volume = {114},
       number = {2},
          eid = {023018},
        pages = {023018},
          doi = {10.1103/8pb2-9t2x},
archivePrefix = {arXiv},
       eprint = {2604.23503},
 primaryClass = {astro-ph.HE},
       adsurl = {https://ui.adsabs.harvard.edu/abs/2026PhRvD.114b3018C}
}

@article{harten1983upstream,
author = {Harten, A. and Lax, P. and Leer, B.},
title = {On Upstream Differencing and Godunov-Type Schemes for Hyperbolic Conservation Laws},
journal = {SIAM Review},
volume = {25},
number = {1},
pages = {35-61},
year = {1983},
doi = {10.1137/1025002},
URL = {https://doi.org/10.1137/1025002},
eprint = {https://doi.org/10.1137/1025002}
}

@ARTICLE{2015JCoPh.286..172C,
       author = {{Cavaglieri}, Daniele and {Bewley}, Thomas},
        title = "{Low-storage implicit/explicit Runge-Kutta schemes for the simulation of stiff high-dimensional ODE systems}",
      journal = {Journal of Computational Physics},
         year = 2015,
        month = apr,
       volume = {286},
        pages = {172-193},
          doi = {10.1016/j.jcp.2015.01.031},
       adsurl = {https://ui.adsabs.harvard.edu/abs/2015JCoPh.286..172C}
}

@ARTICLE{1984JCoPh..54..174C,
       author = {{Colella}, P. and {Woodward}, Paul R.},
        title = "{The Piecewise Parabolic Method (PPM) for Gas-Dynamical Simulations}",
      journal = {Journal of Computational Physics},
         year = 1984,
        month = sep,
       volume = {54},
        pages = {174-201},
          doi = {10.1016/0021-9991(84)90143-8},
       adsurl = {https://ui.adsabs.harvard.edu/abs/1984JCoPh..54..174C}
}

@ARTICLE{1988ApJ...332..659E,
       author = {{Evans}, Charles R. and {Hawley}, John F.},
        title = "{Simulation of Magnetohydrodynamic Flows: A Constrained Transport Model}",
      journal = {\apj},
         year = 1988,
        month = sep,
       volume = {332},
        pages = {659},
          doi = {10.1086/166684},
       adsurl = {https://ui.adsabs.harvard.edu/abs/1988ApJ...332..659E}
}

@ARTICLE{2020PhRvD.102l3015B,
       author = {{Betranhandy}, Aurore and {O'Connor}, Evan},
        title = "{Impact of neutrino pair-production rates in core-collapse supernovae}",
      journal = {\prd},
         year = 2020,
        month = dec,
       volume = {102},
       number = {12},
          eid = {123015},
        pages = {123015},
          doi = {10.1103/PhysRevD.102.123015},
archivePrefix = {arXiv},
       eprint = {2010.02261},
 primaryClass = {astro-ph.HE},
       adsurl = {https://ui.adsabs.harvard.edu/abs/2020PhRvD.102l3015B}
}

@ARTICLE{2000ARA&A..38..113T,
       author = {{Taam}, Ronald E. and {Sandquist}, Eric L.},
        title = "{Common Envelope Evolution of Massive Binary Stars}",
      journal = {\araa},
         year = 2000,
        month = jan,
       volume = {38},
        pages = {113-141},
          doi = {10.1146/annurev.astro.38.1.113},
       adsurl = {https://ui.adsabs.harvard.edu/abs/2000ARA&A..38..113T}
}

@ARTICLE{2023LRCA....9....2R,
       author = {{R{\"o}pke}, Friedrich K. and {De Marco}, Orsola},
        title = "{Simulations of common-envelope evolution in binary stellar systems: physical models and numerical techniques}",
      journal = {Living Reviews in Computational Astrophysics},
         year = 2023,
        month = dec,
       volume = {9},
       number = {1},
          eid = {2},
        pages = {2},
          doi = {10.1007/s41115-023-00017-x},
archivePrefix = {arXiv},
       eprint = {2212.07308},
 primaryClass = {astro-ph.SR},
       adsurl = {https://ui.adsabs.harvard.edu/abs/2023LRCA....9....2R}
}

@BOOK{2020cee..book.....I,
       author = {{Ivanova}, Natalia and {Justham}, Stephen and {Ricker}, Paul},
        title = "{Common Envelope Evolution}",
         year = 2020,
          doi = {10.1088/2514-3433/abb6f0},
       adsurl = {https://ui.adsabs.harvard.edu/abs/2020cee..book.....I}
}

@ARTICLE{2013A&ARv..21...59I,
       author = {{Ivanova}, N. and {Justham}, S. and {Chen}, X. and {De Marco}, O. and {Fryer}, C.~L. and {Gaburov}, E. and {Ge}, H. and {Glebbeek}, E. and {Han}, Z. and {Li}, X.-D. and {Lu}, G. and {Marsh}, T. and {Podsiadlowski}, P. and {Potter}, A. and {Soker}, N. and {Taam}, R. and {Tauris}, T.~M. and {van den Heuvel}, E.~P.~J. and {Webbink}, R.~F.},
        title = "{Common envelope evolution: where we stand and how we can move forward}",
      journal = {\aapr},
         year = 2013,
        month = feb,
       volume = {21},
          eid = {59},
        pages = {59},
          doi = {10.1007/s00159-013-0059-2},
archivePrefix = {arXiv},
       eprint = {1209.4302},
 primaryClass = {astro-ph.HE},
       adsurl = {https://ui.adsabs.harvard.edu/abs/2013A&ARv..21...59I}
}

@ARTICLE{2025Ap&SS.370...11G,
       author = {{Grichener}, Aldana},
        title = "{Mergers of compact objects with cores of massive stars: evolutionary pathways, r-process nucleosynthesis and multi-messenger signatures}",
      journal = {\apss},
         year = 2025,
        month = feb,
       volume = {370},
       number = {2},
          eid = {11},
        pages = {11},
          doi = {10.1007/s10509-025-04402-1},
archivePrefix = {arXiv},
       eprint = {2410.18813},
 primaryClass = {astro-ph.HE},
       adsurl = {https://ui.adsabs.harvard.edu/abs/2025Ap&SS.370...11G}
}

@ARTICLE{1978ApJ...222..269T,
       author = {{Taam}, R.~E. and {Bodenheimer}, P. and {Ostriker}, J.~P.},
        title = "{Double core evolution. I. A 16 M sun star with a 1 M sun neutron-star companion.}",
      journal = {\apj},
         year = 1978,
        month = may,
       volume = {222},
        pages = {269-280},
          doi = {10.1086/156142},
       adsurl = {https://ui.adsabs.harvard.edu/abs/1978ApJ...222..269T}
}

@ARTICLE{1992ApJ...389..546B,
       author = {{Benz}, W. and {Hills}, J.~G.},
        title = "{Three-dimensional Hydrodynamical Simulations of Colliding Stars. III. Collisions and Tidal Captures of Unequal-Mass Main-Sequence Stars}",
      journal = {\apj},
         year = 1992,
        month = apr,
       volume = {389},
        pages = {546},
          doi = {10.1086/171230},
       adsurl = {https://ui.adsabs.harvard.edu/abs/1992ApJ...389..546B}
}

@ARTICLE{1994ApJ...423L..19L,
       author = {{Leonard}, Peter J.~T. and {Hills}, Jack G. and {Dewey}, Rachel J.},
        title = "{A New Way to Make Thorne-Zytkow Objects}",
      journal = {\apjl},
         year = 1994,
        month = mar,
       volume = {423},
        pages = {L19},
          doi = {10.1086/187225},
       adsurl = {https://ui.adsabs.harvard.edu/abs/1994ApJ...423L..19L}
}

@ARTICLE{1993ApJ...411L..33C,
       author = {{Chevalier}, Roger A.},
        title = "{Neutron Star Accretion in a Stellar Envelope}",
      journal = {\apjl},
         year = 1993,
        month = jul,
       volume = {411},
        pages = {L33},
          doi = {10.1086/186905},
       adsurl = {https://ui.adsabs.harvard.edu/abs/1993ApJ...411L..33C}
}

@ARTICLE{1995PhR...256...95C,
       author = {{Chevalier}, R.~A.},
        title = "{Neutron star accretion in dense environments.}",
      journal = {\physrep},
         year = 1995,
        month = may,
       volume = {256},
       number = {1},
        pages = {95-108},
          doi = {10.1016/0370-1573(94)00103-A},
       adsurl = {https://ui.adsabs.harvard.edu/abs/1995PhR...256...95C}
}

@ARTICLE{1996ApJ...459..322C,
       author = {{Chevalier}, Roger A.},
        title = "{Neutrino-cooled Accretion: Rotation and Stellar Equation of State}",
      journal = {\apj},
         year = 1996,
        month = mar,
       volume = {459},
        pages = {322},
          doi = {10.1086/176895},
       adsurl = {https://ui.adsabs.harvard.edu/abs/1996ApJ...459..322C}
}

@ARTICLE{1996ApJ...460..801F,
       author = {{Fryer}, Chris L. and {Benz}, Willy and {Herant}, Marc},
        title = "{The Dynamics and Outcomes of Rapid Infall onto Neutron Stars}",
      journal = {\apj},
         year = 1996,
        month = apr,
       volume = {460},
        pages = {801},
          doi = {10.1086/177011},
archivePrefix = {arXiv},
       eprint = {astro-ph/9509144},
 primaryClass = {astro-ph},
       adsurl = {https://ui.adsabs.harvard.edu/abs/1996ApJ...460..801F}
}

@ARTICLE{2015MNRAS.449..288P,
       author = {{Papish}, Oded and {Soker}, Noam and {Bukay}, Inbal},
        title = "{Ejecting the envelope of red supergiant stars with jets launched by an inspiralling neutron star}",
      journal = {\mnras},
         year = 2015,
        month = may,
       volume = {449},
       number = {1},
        pages = {288-295},
          doi = {10.1093/mnras/stv345},
       adsurl = {https://ui.adsabs.harvard.edu/abs/2015MNRAS.449..288P}
}

@ARTICLE{2019MNRAS.484.4972S,
       author = {{Soker}, Noam and {Grichener}, Aldana and {Gilkis}, Avishai},
        title = "{Diversity of common envelope jets supernovae and the fast transient AT2018cow}",
      journal = {\mnras},
         year = 2019,
        month = apr,
       volume = {484},
       number = {4},
        pages = {4972-4979},
          doi = {10.1093/mnras/stz364},
archivePrefix = {arXiv},
       eprint = {1811.11106},
 primaryClass = {astro-ph.HE},
       adsurl = {https://ui.adsabs.harvard.edu/abs/2019MNRAS.484.4972S}
}

@ARTICLE{2022MNRAS.514.3212H,
       author = {{Hillel}, Shlomi and {Schreier}, Ron and {Soker}, Noam},
        title = "{Three-dimensional simulations of the jet feedback mechanism in common envelope jets supernovae}",
      journal = {\mnras},
         year = 2022,
        month = aug,
       volume = {514},
       number = {3},
        pages = {3212-3221},
          doi = {10.1093/mnras/stac1341},
archivePrefix = {arXiv},
       eprint = {2112.01459},
 primaryClass = {astro-ph.HE},
       adsurl = {https://ui.adsabs.harvard.edu/abs/2022MNRAS.514.3212H}
}

@ARTICLE{2015ApJ...798L..19M,
       author = {{MacLeod}, Morgan and {Ramirez-Ruiz}, Enrico},
        title = "{On the Accretion-fed Growth of Neutron Stars during Common Envelope}",
      journal = {\apjl},
         year = 2015,
        month = jan,
       volume = {798},
       number = {1},
          eid = {L19},
        pages = {L19},
          doi = {10.1088/2041-8205/798/1/L19},
archivePrefix = {arXiv},
       eprint = {1410.5421},
 primaryClass = {astro-ph.SR},
       adsurl = {https://ui.adsabs.harvard.edu/abs/2015ApJ...798L..19M}
}

@ARTICLE{2025ApJ...987...71C,
       author = {{Combi}, Luciano and {Thompson}, Christopher and {Siegel}, Daniel M. and {Philippov}, Alexander and {Ripperda}, Bart},
        title = "{Magnetized Accretion onto Neutron Stars: From Photon-trapped to Neutrino-cooled Flows}",
      journal = {\apj},
         year = 2025,
        month = jul,
       volume = {987},
       number = {1},
          eid = {71},
        pages = {71},
          doi = {10.3847/1538-4357/addc57},
archivePrefix = {arXiv},
       eprint = {2505.00300},
 primaryClass = {astro-ph.HE},
       adsurl = {https://ui.adsabs.harvard.edu/abs/2025ApJ...987...71C}
}

@ARTICLE{2024ApJ...977..196H,
       author = {{Hutchinson-Smith}, Tenley and {Everson}, Rosa Wallace and {Twum}, Angela A. and {Batta}, Aldo and {Yarza}, Ricardo and {Law-Smith}, Jamie A.~P. and {Vigna-G{\'o}mez}, Alejandro and {Ramirez-Ruiz}, Enrico},
        title = "{Rethinking Thorne-{\.Z}ytkow Object Formation: The Fate of X-Ray Binary LMC X-4 and Implications for Ultra-long Gamma-Ray Bursts}",
      journal = {\apj},
         year = 2024,
        month = dec,
       volume = {977},
       number = {2},
          eid = {196},
        pages = {196},
          doi = {10.3847/1538-4357/ad88f3},
archivePrefix = {arXiv},
       eprint = {2311.06741},
 primaryClass = {astro-ph.HE},
       adsurl = {https://ui.adsabs.harvard.edu/abs/2024ApJ...977..196H}
}

@ARTICLE{2024ApJ...971..132E,
       author = {{Everson}, Rosa Wallace and {Hutchinson-Smith}, Tenley and {Vigna-G{\'o}mez}, Alejandro and {Ramirez-Ruiz}, Enrico},
        title = "{Rethinking Thorne-{\.Z}ytkow Object Formation: Assembly via Common Envelope in Field Binaries}",
      journal = {\apj},
         year = 2024,
        month = aug,
       volume = {971},
       number = {2},
          eid = {132},
        pages = {132},
          doi = {10.3847/1538-4357/ad595e},
archivePrefix = {arXiv},
       eprint = {2310.08658},
 primaryClass = {astro-ph.HE},
       adsurl = {https://ui.adsabs.harvard.edu/abs/2024ApJ...971..132E}
}

@ARTICLE{2026ApJ...997...88A,
       author = {{Anninos}, Peter and {Portman}, Matthew E. and {Carmichael}, Scott R. and {Hoffman}, Robert D. and {Sieverding}, Andre},
        title = "{r-process Nucleosynthesis from Hyperaccreting Neutron Stars in Common Envelopes}",
      journal = {\apj},
         year = 2026,
        month = jan,
       volume = {997},
       number = {1},
          eid = {88},
        pages = {88},
          doi = {10.3847/1538-4357/ae2311},
archivePrefix = {arXiv},
       eprint = {2511.18584},
 primaryClass = {astro-ph.HE},
       adsurl = {https://ui.adsabs.harvard.edu/abs/2026ApJ...997...88A}
}

@ARTICLE{2025ApJ...993...61W,
       author = {{Williams}, Lauryn E. and {Chang}, Philip and {Levesque}, Emily M. and {Quinn}, Thomas R.},
        title = "{Neutron Star-Main Sequence Collisions Robustly Form Dynamically Stable Thorne-{\.Z}ytkow Objects}",
      journal = {\apj},
         year = 2025,
        month = nov,
       volume = {993},
       number = {1},
          eid = {61},
        pages = {61},
          doi = {10.3847/1538-4357/ae100d},
archivePrefix = {arXiv},
       eprint = {2510.16129},
 primaryClass = {astro-ph.SR},
       adsurl = {https://ui.adsabs.harvard.edu/abs/2025ApJ...993...61W}
}

@ARTICLE{2025ApJ...978L..38C,
       author = {{Cheong}, Patrick Chi-Kit and {Pitik}, Tetyana and {Longo Micchi}, Lu{\'\i}s Felipe and {Radice}, David},
        title = "{Gamma-Ray Bursts and Kilonovae from the Accretion-induced Collapse of White Dwarfs}",
      journal = {\apjl},
         year = 2025,
        month = jan,
       volume = {978},
       number = {2},
          eid = {L38},
        pages = {L38},
          doi = {10.3847/2041-8213/ada1cc},
archivePrefix = {arXiv},
       eprint = {2410.10938},
 primaryClass = {astro-ph.HE},
       adsurl = {https://ui.adsabs.harvard.edu/abs/2025ApJ...978L..38C}
}

@ARTICLE{2026arXiv260715390S,
       author = {{Sneppen}, Albert and {Hotokezaka}, Kenta and {Irwin}, Christopher M.},
        title = "{Luminous Red Novae as shock-powered transients I: Electron-scattering wings and deviations from Case B}",
      journal = {arXiv e-prints},
         year = 2026,
        month = jul,
          eid = {arXiv:2607.15390},
        pages = {arXiv:2607.15390},
          doi = {10.48550/arXiv.2607.15390},
archivePrefix = {arXiv},
       eprint = {2607.15390},
 primaryClass = {astro-ph.SR},
       adsurl = {https://ui.adsabs.harvard.edu/abs/2026arXiv260715390S}
}

@ARTICLE{2026arXiv260626495M,
       author = {{Mu}, Chunliang and {De Marco}, Orsola and {Berm{\'u}dez-Bustamante}, Luis C. and {Hirai}, Ryosuke and {Price}, Daniel J. and {Siess}, Lionel and {Gonz{\'a}lez-Bol{\'\i}var}, Miguel and {Lau}, Mike Y.~M. and {Blagorodnova}, Nadejda},
        title = "{Dust Formation in Common Envelope Binary Interactions -- III. Lightcurves}",
      journal = {arXiv e-prints},
         year = 2026,
        month = jun,
          eid = {arXiv:2606.26495},
        pages = {arXiv:2606.26495},
          doi = {10.48550/arXiv.2606.26495},
archivePrefix = {arXiv},
       eprint = {2606.26495},
 primaryClass = {astro-ph.SR},
       adsurl = {https://ui.adsabs.harvard.edu/abs/2026arXiv260626495M}
}

@ARTICLE{2025ApJ...994L..41K,
       author = {{Kirilov}, Anthony and {Calder{\'o}n}, Diego and {Pejcha}, Ond{\v{r}}ej and {Duffell}, Paul C.},
        title = "{Two-dimensional Radiation-hydrodynamic Simulations of Luminous Red Novae}",
      journal = {\apjl},
         year = 2025,
        month = dec,
       volume = {994},
       number = {2},
          eid = {L41},
        pages = {L41},
          doi = {10.3847/2041-8213/ae1ae7},
archivePrefix = {arXiv},
       eprint = {2508.09257},
 primaryClass = {astro-ph.SR},
       adsurl = {https://ui.adsabs.harvard.edu/abs/2025ApJ...994L..41K}
}

@ARTICLE{2025ApJ...982...83H,
       author = {{Hatfull}, Roger W.~M. and {Ivanova}, Natalia},
        title = "{Simulating a Stellar Binary Merger. II. Obtaining a Light Curve}",
      journal = {\apj},
         year = 2025,
        month = apr,
       volume = {982},
       number = {2},
          eid = {83},
        pages = {83},
          doi = {10.3847/1538-4357/ada6b8},
archivePrefix = {arXiv},
       eprint = {2412.06583},
 primaryClass = {astro-ph.SR},
       adsurl = {https://ui.adsabs.harvard.edu/abs/2025ApJ...982...83H}
}

@ARTICLE{2021MNRAS.508.2386S,
       author = {{Schreier}, Ron and {Hillel}, Shlomi and {Shiber}, Sagiv and {Soker}, Noam},
        title = "{Simulating highly eccentric common envelope jet supernova impostors}",
      journal = {\mnras},
         year = 2021,
        month = dec,
       volume = {508},
       number = {2},
        pages = {2386-2398},
          doi = {10.1093/mnras/stab2687},
archivePrefix = {arXiv},
       eprint = {2106.11601},
 primaryClass = {astro-ph.HE},
       adsurl = {https://ui.adsabs.harvard.edu/abs/2021MNRAS.508.2386S}
}

@ARTICLE{2017ApJ...834..107B,
       author = {{Blagorodnova}, N. and {Kotak}, R. and {Polshaw}, J. and {Kasliwal}, M.~M. and {Cao}, Y. and {Cody}, A.~M. and {Doran}, G.~B. and {Elias-Rosa}, N. and {Fraser}, M. and {Fremling}, C. and {Gonzalez-Fernandez}, C. and {Harmanen}, J. and {Jencson}, J. and {Kankare}, E. and {Kudritzki}, R.-P. and {Kulkarni}, S.~R. and {Magnier}, E. and {Manulis}, I. and {Masci}, F.~J. and {Mattila}, S. and {Nugent}, P. and {Ochner}, P. and {Pastorello}, A. and {Reynolds}, T. and {Smith}, K. and {Sollerman}, J. and {Taddia}, F. and {Terreran}, G. and {Tomasella}, L. and {Turatto}, M. and {Vreeswijk}, P.~M. and {Wozniak}, P. and {Zaggia}, S.},
        title = "{Common Envelope Ejection for a Luminous Red Nova in M101}",
      journal = {\apj},
         year = 2017,
        month = jan,
       volume = {834},
       number = {2},
          eid = {107},
        pages = {107},
          doi = {10.3847/1538-4357/834/2/107},
archivePrefix = {arXiv},
       eprint = {1607.08248},
 primaryClass = {astro-ph.SR},
       adsurl = {https://ui.adsabs.harvard.edu/abs/2017ApJ...834..107B}
}

@ARTICLE{2026arXiv260300820F,
       author = {{Fryer}, Chris L. and {Burns}, Eric and {Colosimo}, Joseph M. and {Negro}, Michela and {O'Connor}, Brendan},
        title = "{High-Energy Shock Breakout from Supernovae and Gamma-ray Bursts}",
      journal = {arXiv e-prints},
         year = 2026,
        month = feb,
          eid = {arXiv:2603.00820},
        pages = {arXiv:2603.00820},
          doi = {10.48550/arXiv.2603.00820},
archivePrefix = {arXiv},
       eprint = {2603.00820},
 primaryClass = {astro-ph.HE},
       adsurl = {https://ui.adsabs.harvard.edu/abs/2026arXiv260300820F}
}

@ARTICLE{2025ApJ...994..259N,
       author = {{Niblett}, Annabelle E. and {Fryer}, Daniel A. and {Fryer}, Christopher L.},
        title = "{Studying the Power Sources behind Type Ic Supernovae}",
      journal = {\apj},
         year = 2025,
        month = dec,
       volume = {994},
       number = {2},
          eid = {259},
        pages = {259},
          doi = {10.3847/1538-4357/ae1582},
archivePrefix = {arXiv},
       eprint = {2501.15702},
 primaryClass = {astro-ph.HE},
       adsurl = {https://ui.adsabs.harvard.edu/abs/2025ApJ...994..259N}
}

@ARTICLE{2025PhRvD.111l3017S,
       author = {{Shibata}, Masaru and {Fujibayashi}, Sho and {Wanajo}, Shinya and {Ioka}, Kunihito and {Lam}, Alan Tsz-Lok and {Sekiguchi}, Yuichiro},
        title = "{Self-consistent scenario for jet and stellar explosions in collapsar: General relativistic magnetohydrodynamics simulation with a dynamo}",
      journal = {\prd},
         year = 2025,
        month = jun,
       volume = {111},
       number = {12},
          eid = {123017},
        pages = {123017},
          doi = {10.1103/msy2-fwhx},
archivePrefix = {arXiv},
       eprint = {2502.02077},
 primaryClass = {astro-ph.HE},
       adsurl = {https://ui.adsabs.harvard.edu/abs/2025PhRvD.111l3017S}
}

@ARTICLE{2026arXiv260713023H,
       author = {{Hillel}, Shlomi and {Schreier}, Ron and {Soker}, Noam},
        title = "{Simulating the convection in red super-giant stars: wobbling jets in common envelope evolution}",
      journal = {arXiv e-prints},
         year = 2026,
        month = jul,
          eid = {arXiv:2607.13023},
        pages = {arXiv:2607.13023},
          doi = {10.48550/arXiv.2607.13023},
archivePrefix = {arXiv},
       eprint = {2607.13023},
 primaryClass = {astro-ph.SR},
       adsurl = {https://ui.adsabs.harvard.edu/abs/2026arXiv260713023H}
}

@ARTICLE{2026arXiv260710267G,
       author = {{Gurjar}, Deepanshu and {Chamandy}, Luke and {Blackman}, Eric G. and {Zou}, Yangyuxin and {Liu}, Baowei and {Nordhaus}, Jason},
        title = "{Effect of Neutron Star Jets on Common Envelope Evolution}",
      journal = {arXiv e-prints},
         year = 2026,
        month = jul,
          eid = {arXiv:2607.10267},
        pages = {arXiv:2607.10267},
          doi = {10.48550/arXiv.2607.10267},
archivePrefix = {arXiv},
       eprint = {2607.10267},
 primaryClass = {astro-ph.SR},
       adsurl = {https://ui.adsabs.harvard.edu/abs/2026arXiv260710267G}
}

@ARTICLE{2025RAA....25b5023S,
       author = {{Soker}, Noam},
        title = "{Jets are the Most Robust Observable Ingredient of Common Envelope Evolution}",
      journal = {Research in Astronomy and Astrophysics},
         year = 2025,
        month = feb,
       volume = {25},
       number = {2},
          eid = {025023},
        pages = {025023},
          doi = {10.1088/1674-4527/adb15b},
archivePrefix = {arXiv},
       eprint = {2412.04017},
 primaryClass = {astro-ph.SR},
       adsurl = {https://ui.adsabs.harvard.edu/abs/2025RAA....25b5023S}
}

@ARTICLE{2017MNRAS.470.2929M,
       author = {{Moreno M{\'e}ndez}, Enrique and {L{\'o}pez-C{\'a}mara}, Diego and {De Colle}, Fabio},
        title = "{Dynamics of jets during the common-envelope phase}",
      journal = {\mnras},
         year = 2017,
        month = sep,
       volume = {470},
       number = {3},
        pages = {2929-2937},
          doi = {10.1093/mnras/stx1385},
archivePrefix = {arXiv},
       eprint = {1702.03293},
 primaryClass = {astro-ph.HE},
       adsurl = {https://ui.adsabs.harvard.edu/abs/2017MNRAS.470.2929M}
}

@ARTICLE{2019MNRAS.482.3646L,
       author = {{L{\'o}pez-C{\'a}mara}, Diego and {De Colle}, Fabio and {Moreno M{\'e}ndez}, Enrique},
        title = "{Self-regulating jets during the common-envelope phase}",
      journal = {\mnras},
         year = 2019,
        month = jan,
       volume = {482},
       number = {3},
        pages = {3646-3655},
          doi = {10.1093/mnras/sty2959},
archivePrefix = {arXiv},
       eprint = {1806.11115},
 primaryClass = {astro-ph.HE},
       adsurl = {https://ui.adsabs.harvard.edu/abs/2019MNRAS.482.3646L}
}

@ARTICLE{2020MNRAS.497.2057L,
       author = {{L{\'o}pez-C{\'a}mara}, Diego and {Moreno M{\'e}ndez}, Enrique and {De Colle}, Fabio},
        title = "{Disc formation and jet inclination effects in common envelopes}",
      journal = {\mnras},
         year = 2020,
        month = sep,
       volume = {497},
       number = {2},
        pages = {2057-2065},
          doi = {10.1093/mnras/staa1983},
archivePrefix = {arXiv},
       eprint = {2004.04158},
 primaryClass = {astro-ph.HE},
       adsurl = {https://ui.adsabs.harvard.edu/abs/2020MNRAS.497.2057L}
}

@ARTICLE{2026MNRAS.551g1542S,
       author = {{Sakurai}, Daiyu and {Akaho}, Ryuichiro and {Yamada}, Shoichi},
        title = "{Numerical studies of accretion flows on to a neutron star engulfed in a massive star}",
      journal = {\mnras},
         year = 2026,
        month = sep,
       volume = {551},
       number = {3},
          eid = {stag1542},
        pages = {stag1542},
          doi = {10.1093/mnras/stag1542},
archivePrefix = {arXiv},
       eprint = {2604.19236},
 primaryClass = {astro-ph.HE},
       adsurl = {https://ui.adsabs.harvard.edu/abs/2026MNRAS.551g1542S}
}

\end{document}